\documentclass[aps,prd,10pt,notitlepage,nofootinbib,superscriptaddress,showkeys,showpacs]
{revtex4-1}

\usepackage{amsfonts, amsmath, amssymb, amsthm}

\usepackage{color}
\usepackage{stmaryrd}
\usepackage{graphicx, array}
\newcolumntype{M}[1]{>{\centering\arraybackslash}m{#1}}
\usepackage{amsmath}
\usepackage{bbm}
\usepackage{hyperref}

\newcommand{\be}{\begin{equation}}
\newcommand{\ee}{\end{equation}}
\newcommand{\bea}{\begin{eqnarray}}
\newcommand{\eea}{\end{eqnarray}}
\renewcommand\[{\begin{equation}}
\renewcommand\]{\end{equation}}
\renewcommand{\eqref}[1]{Eq.\,(\ref{#1})}

\newcommand{\cA}{\mathcal{A}}
\newcommand{\cM}{\mathcal{M}}

\newcommand{\cB}{\mathcal{B}}
\newcommand{\cF}{\mathcal{F}}

\newcommand{\cG}{\mathcal{G}}

\newcommand{\cO}{\mathcal{O}}
\newcommand{\cT}{\mathcal{T}}
\newcommand{\cP}{\mathcal{P}}

\newcommand{\cI}{\mathcal{I}}

\newcommand{\cZ}{\mathcal{Z}}

\newcommand{\gG}{\Gamma }

\newcommand{\bB}{\mathbb{B}}
\newcommand{\bC}{\mathbb{C}}

\newcommand{\un}{\mathbbm{1}}

\newcommand{\Tr}{\mathrm{Tr}}

\def\ie{{i.e.}~}
\def\eg{{e.g.}~}

\def\d{\mathrm{d}}
\def\e{\textrm e}

\newcommand{\gn}{G_{\textsc{n}}}
\newcommand{\rk}{D}		

\newcommand{\ktt}{K_{3,3}}

\renewcommand{\sb}{{s_\cB}}
\newcommand{\bmp}[1]{\mathcal{B}_{{#1}_{\textsc{M}}}}
\newcommand{\bnp}[1]{\mathcal{B}_{{#1}_{\textsc{N}}}}
\newcommand{\bk}{\mathcal{B}_{3_{\textsc{K}}}}
\newcommand{\tmp}[1]{t_{{#1}_{\textsc{M}}}}
\newcommand{\tnp}[1]{t_{{#1}_{\textsc{N}}}}
\newcommand{\tm}[1]{t_{#1}}
\newcommand{\tn}[1]{g_{#1}}
\newcommand{\tk}{t_{{3_\textrm{\sc k}}}}
\newcommand{\tp}{g}
\newcommand{\tc}{t_\text{cr}}
\newcommand{\tra}{{\text{pt}}}
\newcommand{\cri}{{\text{cr}}}
\newcommand{\pla}{{\text{pl}}}
\newcommand{\bp}{{\text{bp}}}

\newcommand{\gt}{G_2} 
\newcommand{\lop}{{V}}       

\newcommand{\lm}[1]{\lambda_{#1}}
\newcommand{\lp}{\lambda}
\newcommand{\lopt}{{\tilde{V}}}  

\newcommand{\nsgf}{\cP_\textrm{ns}}

\newcommand{\B}{B}
\newcommand{\para}[1]{

\

\noindent{\it #1}

\

}

\begin{document}

\title{Multi-critical behaviour of 4-dimensional tensor models up to order 6}

\author{Luca Lionni}
\email{luca.lionni@th.u-psud.fr}
\affiliation{Laboratoire de Physique Th\'eorique, CNRS UMR 8627, Universit\'e Paris XI, 91405 Orsay Cedex, France}
\affiliation{LIPN, UMR CNRS 7030, Institut Galil\'ee, Universit\'e Paris 13, Sorbonne Paris Cit\'e, 99, avenue Jean-Baptiste Cl\'ement, 93430 Villetaneuse, France}

\author{Johannes Th\"urigen}
\email{johannes.thuerigen@physik.hu-berlin.de}
\affiliation{Departments of Physics and Mathematics,  Humboldt University,  Unter den Linden 6, 10099 Berlin, Germany}
\affiliation{Laboratoire de Physique Th\'eorique, CNRS UMR 8627, Universit\'e Paris XI, 91405 Orsay Cedex, France}

\date{\today}

\begin{abstract} 
Tensor models generalize the matrix-model approach to 2-dimensional quantum gravity to higher dimensions. 
Some models allowing a $1/N$ expansion have been explored, most of them generating branched-polymer geometries. 
Recently, enhancements yielding an additional 2d pure gravity (planar) phase and an intermediate regime of proliferating baby universes have been found. 
It remains an open issue to find models 
escaping these lower-dimensionality universality classes.

Here we analyse the dominant regime and critical behaviour of a range of new models which are candidates for such effective geometries,
in particular interactions based on the utility graph $\ktt$. 
We find that, upon proper enhancement, the two-phase structure of a branched-polymer and a 2d gravity regime is the common case in $U(N)$-invariant 
rank $\rk=4$ tensor models of small orders. 
Not only the well known so-called necklace interactions but also $\ktt$-type interactions turn out as the source for the planar regime.
We give a systematic account of the enhancement scaling, the counting of leading-order diagrams and the multi-critical behaviour of a wide range of interactions, in particular for all order-6 interactions of rank 3 and 4.
These findings support the claim of universality of such mixtures of branched-polymer and planar diagrams at criticality. In particular, this hints at the necessity to consider new ingredients, or interactions of higher order and rank, in order to obtain higher dimensional continuum geometry from tensor models. 
\end{abstract}

\maketitle

\section*{Introduction}

Tensor models \cite{Gurau:2016wk,Gurau:2016ed,Gurau:2012hl}, which generalize matrix models \cite{DiFrancesco:1995ih}  to higher dimensions, were introduced as a non-perturbative approach to quantum gravity, and an analytical tool to explore random geometries in dimension higher than two.
These geometries are generated as Feynman diagrams in the perturbative expansion of a tensor path integral,
and the amplitudes associated to discrete geometries are discretizations of the quantum path integral of general relativity.
Restricting the tensor action to be $U(N)^\rk$-invariant such that interactions can be represented as $\rk$-coloured graphs \cite{Bonzom:2012bg} allows for a $1/N$ expansion \cite{Gurau:2011aq,Gurau:2012ek}.
A continuum limit can be obtained by tuning correlation functions to criticality in the coupling parameters \cite{Bonzom:2011cs,Bonzom:2012bg,Bonzom:2013in}.

So far, only continuous geometries which come as limits of one or two dimensional structures have been identified.
Tensors of rank $\rk=1$, \ie vector models, generate plane trees in the large-$N$ limit, which converge towards the continuous random tree \cite{Aldous:1991kr, Aldous:1991ja, Aldous:1993ed}  in the continuum limit, also called branched polymers, and are characterized by the entropy critical exponent $\gamma=1/2$, Hausdorff dimension $d_h=2$ and spectral dimension $d_S=4/3$.
In the $\rk=2$ case of matrix models, amplitudes scale as $N^{2-2g}$ leading to planar geometries (genus $g=0$) in the large-$N$ limit, which converge at criticality towards the Brownian sphere \cite{Marckert:2006fs,LeGall:2008wd,LeGall:2007ew},
with $\gamma=-1/2$, $d_H=4$, and $d_S=2$, which has recently been proven to be equivalent to 2d Liouville quantum gravity \cite{POLYAKOV1981207, Miller:2015tx,Miller:2016tg,Miller:2016ut,David:2016jc}.
Tensor models of rank $\rk\ge3$ are generically dominated in the large-$N$ limit by so-called melonic diagrams \cite{Bonzom:2011cs}
encoding discrete $\rk$-dimensional spherical manifolds. Surprisingly, in the continuum limit, they converge towards branched polymers \cite{Gurau:2013th}. Nevertheless, it has recently been shown that non-melonic interactions coined ``necklaces" can be enhanced to obtain continuum 2d geometry in this limit \cite{Bonzom:2015gt} using an intermediate-field (IF) representation \cite{Gurau:2014kj,Delepouve:2014wk}.


The question is thus whether it is possible to find an enhancement of certain interactions such that the  large-$N$ continuum limit has properties suited for a quantum space-time of dimension higher than two.
Such new continuum geometries should have spectral dimension $d_S>2$ and are expected to be characterized also by new critical exponents $\gamma$. While there are no guiding lines from mathematics where no such geometries are known so far, tensor models generating $\rk$-dimensional discrete topologies could, in principle, contain such new universality classes in the continuum limit.

The motivation of the current work is that there are classes of tensor models not investigated before which, on first sight, have the potential to describe such new universality classes of higher-dimensional continuum geometries.
The naive idea is that 
such geometries might arise when there are sufficiently large non-planarities, \eg crossing edges on the diagrams such that the diagrams cannot be embedded on the 2-sphere. 
Intuitively, a space obtained as a limit of discrete random spaces with diverging number of holes competing with the diverging volume should have interesting fractal properties.
Natural candidates are non-local interactions with a combinatorial structure based on the utility graph $\ktt$.
According to Kuratowski's theorem \cite{Kuratowski:1930ta}, finite graphs are not planar if and only if they contain (a subdivision of) either the utility graph $\ktt$ or the complete graph $K_5$. 
As bipartite graphs, $U(N)^\rk$-invariant tensor interactions cannot contain $K_5$. It is thus natural to expect new quantum geometries to result from sufficiently dominant interactions which are extensions of $\ktt$.

\

To test this hypothesis we have explored the full space of sixth-order tensor-invariant interactions at rank $\rk=3$ and 4.
There are preliminary results that the $\ktt$-interaction for $\rk=3$ leads only to small crossings \cite{SWMaps}, which should not be dominant in the continuum limit, as we indeed prove in this paper.
At $\rk=4$, it is possible to have $\ktt$-type interactions which lead,  on first sight, to large crossings.
Nevertheless, studying the dominant diagrams and the critical behaviour in detail we find that these models are still in the universality class of branched-polymer and 2d-gravity geometries. We find an explanation for this in terms of a bijection between the Feynman diagrams and embedded diagrams, that is, graphs drawn on surfaces.

Our systematic study of non-melonic tensor models covers a number of interesting aspects. 
First of all, 
it is the first study of interactions of order higher than 4,
apart from a specific rank-3 model of order 8 leading to branched polymers in the continuum at large $N$ \cite{Bonzom:2016vc}.
So far, the focus in the literature has been on quartic interactions \cite{Dartois:2013kb, Delepouve:2014wk,Delepouve:2015hc,  Nguyen2015, Delepouve:2016gi, Delepouve:2017} and straightforward extensions.
We are able to extend these studies to higher order using the recent generalization of intermediate-field representation to arbitrary order \cite{SWMaps}. 
This method relies on a bijection between stranded Feynman diagrams and embedded diagrams.
It is a powerful tool for finding enhancements of less dominant interactions and for identifying the resulting diagrams at leading order in the $1/N$ expansion. 
We provide a systematic analysis of the enhancement scalings of all sextic interactions.

The bijection to simpler embedded diagrams provides means to determine amplitudes from counting diagrams.
Tensor models, just like matrix models, are purely combinatorial in that amplitudes in the perturbative expansion are given, up to factors in couplings and $N$, simply by the number of diagrams contributing at that order. 
Accordingly, leading-order correlation functions are generating functions in the combinatorics sense. 
In this way we obtain explicit Dyson-Schwinger-type equations for two-point functions and find solutions solving them.
On these grounds we are able to determine the critical behaviour in many cases by performing explicitly the Newton-Puiseux expansion in fractional powers at critical loci. The critical behaviour of correlations, characterized by the entropy exponent $\gamma$, provides a first hint of what the continuum limit of the model is.

As a consequence, we are for the first time able to analyse models generating topologically non-trivial discrete geometries. 
Indeed, the tensor interactions are dual to $D$-dimensional building blocks, and the Feynman diagrams are dual to discrete spaces obtained by gluing such building blocks together.
So far, only models involving interactions with boundaries of spherical topology, thus generating discrete geometries obtained from gluing balls, have been investigated.
The interactions we are considering in this paper are dual to building blocks whose boundaries are spherical, toroidal, or may even be singular (pseudo)-manifolds. Still, we find that they all belong to the universality classes with continuum limits of branched-polymer or planar geometries.

That is, all models considered here belong to the universality classes of branched polymers ($\gamma=1/2$), of 2d pure gravity (planar diagrams, with $\gamma=-1/2$), or to an intermediate transitional regime of ``proliferating baby universes", with critical exponent $\gamma=1/3$, through various combinations of enhanced interactions.
This last regime was discovered in matrix models by considering modified Einstein-Hilbert actions with higher curvature terms, resulting into multi-trace interactions \cite{Das:1990gp,Korchemsky:1992uj,AlvarezGaume:1992np,Ambjorn:1993vz,Gubser:1994cg}.
On the contrary,
we recover these regimes as different phases of tensor models with connected 
interactions, 
corresponding to discrete theories of gravity without higher-order curvature terms.
In combinatorics, these three regimes appear also in the context of (non-embedded) planar graphs with given 2- or 3-connected components \cite{Gimenez:2012ev}. 

A peculiar feature of tensor models is that the branched-polymer--planar phase diagram can even be obtained from a single type of interaction. Indeed we show here that it occurs already for two quartic necklace interactions of different colouring, extending and completing previous work \cite{Bonzom:2015gt,Bonzom:2016to}. In the present model, the branched-polymer phase is not obtained from a planar regime by coupling it to a polymerization term, as done in \cite{Das:1990gp,Korchemsky:1992uj, AlvarezGaume:1992np, Bonzom:2015gt,Bonzom:2016to}, but rather as an interference between planar regimes, which instead of building up interesting behaviour, reduces to a branched polymer regime for a small region of the phase space. As expected, at the boundary of this critical region we observe the intermediate regime of proliferation of baby universes.

The multi-critical structure of the models turns out to be more involved than anticipated in the literature \cite{Das:1990gp,AlvarezGaume:1992np,Bonzom:2015gt}.
It is well known that with several couplings one can have sequences of multi-critical subspaces with critical exponents $\gamma = 1/2,\, 2/3,\, 3/4,...$ in the branched-polymer regime \cite{Ambjorn:1990ky}, and the Kazakov exponents $\gamma = -1/2,\, -1/3,\, -1/4,...$ in the planar regime \cite{Kazakov:1989gw} which in the continuum comes to coupling non-unitary conformal matter to 2d gravity.
If such multi-critical subspaces are incident with the baby-universe transition regime, special multi-critical values could be found. In the case of matrix models it is stated that such cases occur \cite{Korchemsky:1992uj, AlvarezGaume:1992np}.
However, our explicit calculations of critical loci in the tensor models considered here indicate that branched-polymer-regime  multi-critical subspaces are not incident with the transition regime.
Instead, we find that the presence of couplings relevant for the planar regime can enrich also the multi-critical structure of the branched-polymer regime.

Apart from these new details in the multi-critical structure, our main result is the confirmation that order-6 tensor models of rank 3 and 4 belong still to the same universality classes as order-4 models. Initial hope to find new universality classes for $\ktt$-type interactions could not be fulfilled. 
Indeed, the bijection with embedded diagrams shows clearly that the influence of non-planarities in the interactions on the structure of leading order diagrams is not strong enough by itself to change their large scale properties, and therefore the corresponding continuum limit. 
This could be taken as an argument that $U(N)^D$-invariant tensor models do not cover new regimes.

Conversely, the absence of new regimes in the models considered might suggest that the structure of the embedded interaction, after the intermediate-field bijection, should be a more pertinent criterion than the non-planarity of its stranded version. 
Indeed, all order 4 and 6 tensor model diagrams of rank 3 and 4 allow for some bijection to two-dimensional combinatorial structures.
The non-planarity becomes a condition on colourings of edges of ribbon graphs which at best has no effect with respect to their usual planar large-$N$ behaviour, or tends to prevent large loops from appearing instead of favouring them, as was initially hoped for.  
However, this argument is not applicable in general for higher-order or higher-rank models since the bijection maps interactions to ``stuffed" embedded vertices with a non-trivial internal structure in general. 

\

The structure of the paper is the following: In the first section, we introduce rank-$D$ enhanced random tensor models, describe their relevance as discrete formulations of gravity in dimension $D$, and comment on known critical behaviour and continuum limits. We introduce some of the models we solve in the following. The exhaustive list of order-3 and order-4 $U(N)^D$-invariant interactions can be found in Tables~\ref{table:rank3-interactions} and~\ref{table:rank4-interactions}, as well as the scalings and critical exponents computed in this paper. Section~\ref{sec:LOdiagrams} is dedicated to the characterization of leading order Feynman diagrams throughout the bijection introduced in \cite{SWMaps} with embedded diagrams. 
Finally, in Section~\ref{sec:MultiCrit}, we analyse the critical and multi-critical behaviour of the models.


\section{Tensor models with branched-polymer--planar-transition critical behaviour}
\label{sec:SecI}

\subsection{Tensor models}

We consider tensor models as defined by a partition function 
with a finite number of $U(N)^{\otimes D}$-invariant interactions 
indexed by $\cB$ in the set of interactions $\bB$ defining the model,
\be
\label{eqref:PartFunct}
\cZ_{\bB}\bigl(\{\tm\cB\}, N\bigr)=\int_{\bC^D}\exp \biggl[N^{D-1}\sum_{\cB\in\bB}{\tm\cB}{N^\sb}  \Tr_{\cB}(T,  \bar{T})\biggr]d\mu_{0}(T,  \bar{T}), 
\ee
where $s_\cB\in\mathbb{N}$ are the scalings, and $d\mu_0(T, \bar{ T})$ is the Gaussian measure
\be
\label{eqref:gaussianmeasure}
d\mu_0(T, \bar{ T}) = \frac{1}{\cZ_0} e^{- N^{D-1} T\cdot \bar{ T}}\ \prod_{a_1, \dotsc, a_D=1}^N dT_{a_1 \dotsb a_D} d\bar{T}_{a_1 \dotsb a_D},
\ee
in which we denoted $T\cdot \bar T=\sum_{a_1,\cdots a_D}T_{a_1 \dotsb a_D} \bar{T}_{a_1 \dotsb a_D}$.
The partition function $\cZ_0$ is normalized such that $\cZ_{\bB}(0, N)=1$. 
More specifically, $\Tr_{\cB}(T,  \bar{T})$ are polynomials of $T$ and $\bar T$ invariant under unitary transformations in $U(N)^{\otimes D} = U(N)\otimes \cdots\otimes U(N)$.
Therefore they must have the same number of copies of $T$ and $\bar T$, and indices may only be contracted between a $T$ and a $\bar T$. An example in $D=4$ is the following,
\be 
\includegraphics[scale=1.05]{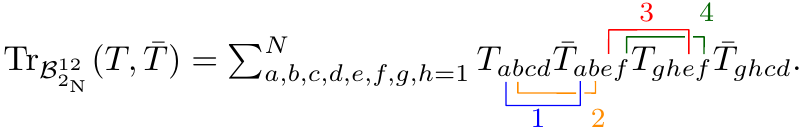}
\ee 

An interaction $\cB$, usually called a \emph{bubble}, is represented graphically by drawing a line labelled with the index $c$ between a $T$ and a $\bar T$ if the index $c$ is summed between the two corresponding tensors, as pictured below on the left for the previous example.
\be 
\label{fig:Neck}
\begin{array}{c}
\includegraphics[scale=0.6]{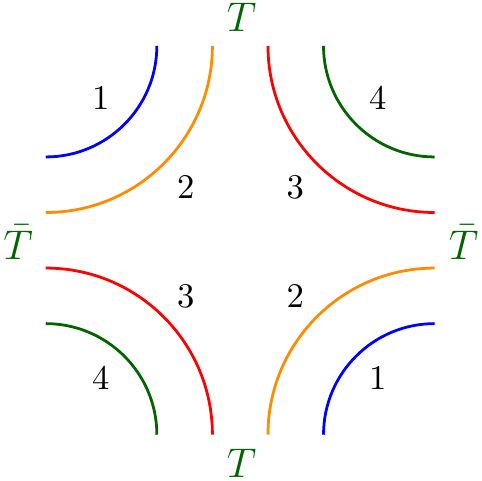}
\end{array}
\hspace{3cm}
\begin{array}{c}\includegraphics[scale=0.6]{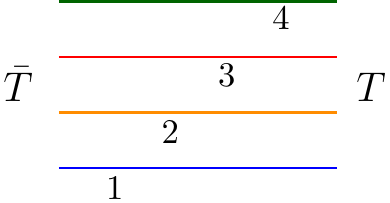}\end{array}
\ee 

For a given choice of set $\bB$ of interactions, the perturbative 
expansion of the free energy 
\be 
\cF_\bB\bigl(\{\tm\cB\},N\bigr)= \ln\bigl(\cZ_\bB(\{\tm\cB\},N)\bigr)
\ee 
is labelled by connected Feynman diagrams which are  generalized ribbon graphs, also called stranded graphs. 
The propagators contain $\rk$ strands, one for each index of $T$, as shown above on the right of (\ref{fig:Neck}), and the interactions $\cB\in\bB$ have a non-trivial internal structure.
\be 
\label{fig:FeynNeck}
\begin{array}{c}\includegraphics[scale=0.4]{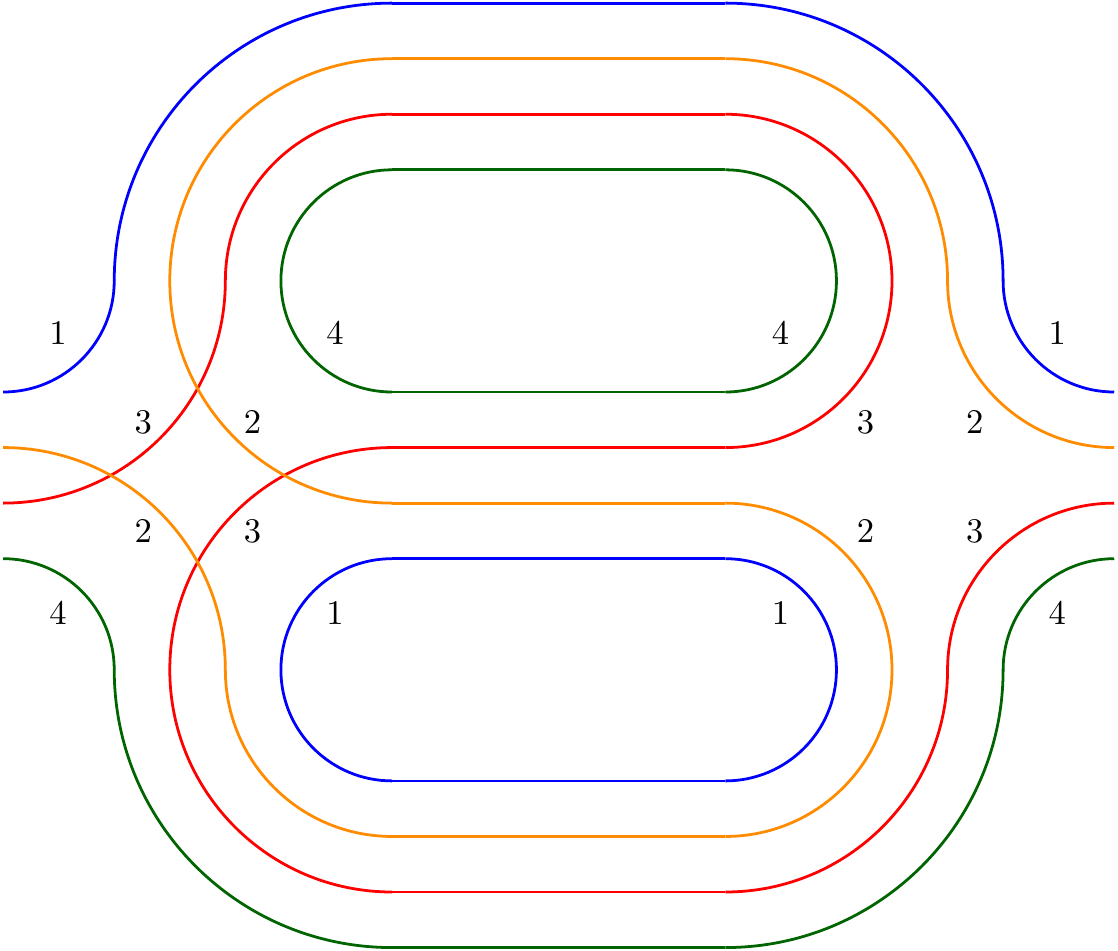}\end{array}
\ee 
Figure~(\ref{fig:FeynNeck}) is an example of a contribution to the 2-point function of a quartic rank-4 model. 
%
The corresponding diagrams $\gG$ contain a certain number $F_c(\gG)$ of closed curves for each colour $c$, which we name {\it faces}. We denote 
\be 
\label{eqref:Faces}
F(\gG)=\sum_{c=1}^D F_c(\gG)
\ee 
the total number of faces of a diagram $\gG$. 
In the following, $n_\cB$ is half the order of the interaction $\cB$. Using the fact that there are $2 n_\cB$ half-propagators per interaction $\cB$, each contributing with a factor $N^{-(D-1)}$, the amplitude associated to a Feynman diagram $\gG$ is  
\be
\label{eqref:amp}
\cA(\gG) = \frac1{c(\Gamma)} \bigl(\prod _{\cB\in\bB}\tm\cB ^{\B_\cB(\gG)}\bigr)N^{\sum_{\cB\in\bB}\B_\cB(\gG)\bigl(s_\cB-(n_\cB-1)(D-1)\bigr) +F(\gG)},
\ee
where $\B_\cB(\gG)$ is the number of interaction bubbles of type $\cB$ in $\gG$,  and $c(\Gamma)$ is a symmetry factor.

The $1/N$ expansion of the free-energy of such a random tensor model can be understood as a discrete version of the formal Einstein-Hilbert partition function of $D$-dimensional pure Euclidean gravity \cite{Gurau:2016ed}, 
\be
\label{EH1}
\cZ_\textsc{EH} = \sum_{\text{topologies}} \int \d g \, \e^{-S_\textsc{EH}} \quad , \quad
S_\textsc{EH} =\frac{1}{16\pi\gn} \int d^Dx\sqrt {\lvert g\rvert} (2\Lambda-R),
\ee
where $\Lambda$ is the cosmological constant, $\gn$ is Newton's constant, and $R$ is the Ricci scalar of the metric $g$. Indeed, every discrete manifold can be equipped with a geometry by taking all the edges to have a fixed length. 
Therefore, the integration over geometries can be replaced with a sum over discretized manifolds of a discrete version of the Einstein-Hilbert action.  
Stranded tensor diagrams $\gG$ are dual to 
pseudo-manifolds 
such that for some set $\bB$ of interactions 
\be
\label{PartFunc1}
\cZ_\textsc{EH}^\bB = \cF_\bB\bigl(\{\tm\cB\},N\bigr) = \sum_{\gG} \e^{\sum_{\cB\in\bB} \B_\cB(\gG)\bigl(\kappa_{D-2}(\gn)F_\text{int}(\cB)-\kappa_D(\gn,\Lambda)2n_\cB\bigr) + \kappa_{D-2}(\gn)F(\gG)  },
\ee
where $\kappa_D(G,\Lambda)$ and $\kappa_{D-2}(G)$ 
do not depend on $\cB$ and are obtained from Regge's action \cite{Regge:1961ct} for fixed lengths, and the number of internal faces $F_\text{int}(\cB)$ depends only on the stranded structure of the interaction $\cB$. We recover the amplitudes\footnote{To get rid of the symmetry factor, we can just consider stranded diagrams with one distinguished propagator.} \eqref{eqref:amp} by setting 
\be
\label{eqref:EquivEHTensor}
N=\e^{\kappa_{D-2}(\gn)}, \qquad\text{ and } \qquad \tm\cB=\e^{\kappa_{D-2}(\gn) \bigl(F_\text{int}(\cB) - s_\cB + (n_\cB-1)(D-1)\bigr) - 2\kappa_D(\gn,\Lambda)n_\cB}.
\ee
This explains the physical meaning of parameters $N$ and $\tm\cB$ in tensor-model quantum gravity, and justifies that enhanced tensor models can be interpreted as discrete Einstein-Hilbert theories of gravity without higher order curvature terms, something which has not been stressed before.
In particular, the large-$N$ limit corresponds to a weak-coupling limit in quantum gravity as $\kappa_{D-2}(\gn)$ is inversely proportional to Newton's constant $\gn$ \cite{Gurau:2016ed}.

\para{Continuum limit and critical behaviour}

\noindent In the large $N$ limit, only contributions which maximize the exponent of $N$ in \eqref{eqref:amp} survive.  This corresponds to the physical limit of small Newton's constant. 
The large-$N$ free energy, 
\be 
\cF_{LO}\bigl(\{\tm\cB\}\bigr)=\lim_{N\rightarrow \infty} \frac 1 {N^D} \cF\bigl(\{\tm\cB\},N\bigr),
\ee 
is a formal power series in the coupling constants. There are regions of the phase space for which $\cF_{LO}$ becomes singular (the inverse functions develop extrema). 
When reaching a singularity, the correlation functions do not depend anymore on the details of the Feynman diagrams but only on the way diagrams with large number of interactions scale. This corresponds to a continuum limit, for which the tensor model could, in some relevant cases, describe a non-perturbative theory of random fractal continuum spaces, a short-scale version of the Einstein-Hilbert action for pure gravity. 

The relevance of such a continuum limit is partially encoded in the critical behaviour of the correlation functions when reaching singularities, that is, by the corresponding entropy critical exponent $\gamma$, also known in matrix models as the \emph{string susceptibility} \cite{DiFrancesco:1995ih}. It is obtained from the smallest non-integer exponent of the power expansion of the correlation functions around a singular point. For instance, in the simpler case of a single coupling $\tm{}$ with dominant singularity $\tm\cri$, the critical exponent $\gamma$ is given by 
\be
\cF_{LO}(\tm{}) = \sum_{i=0}^{k-1} \alpha_i (\tm\cri - \tm{})^{a_i}  + \alpha_k (\tm\cri - \tm{})^{2-\gamma} + o(\tm\cri - \tm{})^{2-\gamma},
\ee
where the $\alpha_i$ may be null, all $a_i$ are integers and $2-\gamma$ is not. In the case of a $2n$-point function the first non-integer exponent is $2-n-\gamma$. In practice, it is easier to compute the critical exponent from the 2-point function since then the symmetry factors $c(\gG)$ of the amplitudes in \eqref{eqref:amp} are trivial. 
We interpret the critical exponent $\gamma$ as follows.
\begin{itemize}
\item If $\gamma = 1/2$, the continuum limit is very likely to describe branched polymers \cite{Gurau:2013th}, also known as the continuous random tree \cite{Aldous:1991kr,Aldous:1991ja,Aldous:1993ed}.
This is the limit reached in random vector models 
or coloured random tensor models \cite{Bonzom:2011cs},
and it is not a relevant limit for quantum gravity in our context\footnote{We stress however that such tensor models can be of interest as Sachdev-Ye-Kitaev (SYK) like tensor models when adding a time parameter \cite{Asimplemodelofqu:2015wt, Witten:2016uz,Gurau:2016lzk,Bonzom:2017gr}. In some cases, it is believed that they could serve as toy models to study the quantum properties of black-holes \cite{Maldacena:2016hq,Polchinski:2016bh}.}. 
We expect branched polymers in the continuum when the Feynman diagrams are in one-to-one correspondence with families of embedded trees which do not carry too much additional information\footnote{In many cases, there is a one-to-one correspondence between planar diagrams and families of plane trees which carry labelings \cite{Cori:1981ev,Schaeffer:1998ta}. This additional information is enough to escape the universality class of branched polymers.}. 
\item If the critical exponent is $\gamma=-1/2$, we are in the universality class of pure 2d gravity, and we expect the continuum limit to be the Liouville theory of 2d quantum gravity \cite{POLYAKOV1981207, Miller:2015tx,Miller:2016tg,Miller:2016ut,David:2016jc}.
Intuitively, this behaviour is expected when there is a one-to-one correspondence between the Feynman diagrams and families of planar diagrams which do not carry too much information, and which are not restricted by criteria so strong they change the universality class, e.g. by restricting the number of possible loops (embedded trees are planar).

\item If the critical exponent is $\gamma=1/3$, and if the characterization of contributing diagrams suggests so, we expect baby universes to proliferate \cite{Das:1990gp,Korchemsky:1992uj,AlvarezGaume:1992np}. 
This balanced regime interpolates between branched polymers and pure 2d gravity. It is obtained in the context of matrix models by adding a non-local\footnote{Non-local refers here to the fact that the stranded interaction is non-connected.} $\Tr(M^2)^2$ polymerization perturbation to the potential. Such a multi-trace term encodes higher order curvature perturbations to the Einstein-Hilbert action and generates nodal points at which the discrete surfaces generated by one-matrix models ``touch" each other without interacting. At leading order, the incidence relations between these surfaces are tree-like. %
If the coupling of the non-local term is too weak, the perturbation has no effect and the universality class is that of pure 2d gravity. If the polymerization is too strong, the nodal points dominate, and branched polymers are recovered in the continuum.

There is an intermediate value of the coupling for which the proliferation of nodal points competes with the diverging areas of the surfaces, leading to an intermediate regime for which baby universes proliferate, and the entropy exponent $\gamma=1/3$ is positive. We stress however that, for matrix models, this regime is reached at the cost of introducing non-local interactions corresponding to higher order curvature terms. 
In this paper, we study various 4-dimensional models with local  interactions\footnote{In the sense that stranded interactions are connected.}.
Furthermore, they correspond to $D$-dimensional  Einstein-Hilbert Euclidean partition functions without higher order curvature terms throughout relations \eqref{eqref:EquivEHTensor}. Still, they entail the same three regimes.
\end{itemize}
%

\para{Multi-critical behaviour}

For some regions of the phase space, the correlation functions exhibit more singular behaviour. For instance, this occurs when higher derivatives of the inverse functions vanish. To reach such regions, one typically has to fine-tune the couplings. Higher order critical exponents appear,  which depend on how singular the correlation functions get, but also on the critical exponents when the couplings are not fine-tuned. For instance, when the generic critical 
behaviour is that of branched polymers $\gamma=1/2$, the  multi-critical exponent of order $m$ is \cite{Ambjorn:1990ky}
\be
\gamma=\frac{m}{m+1}.
\ee
When the generic critical 
behaviour is that of 2d pure gravity, the order-$n$ multi-critical exponent is \cite{Kazakov:1989gw}
\be
\gamma= - \frac1{n+1}.
\ee
To obtain a multi-critical exponent of order $q=m$ (branched polymer) or $q=n$ (planar), one typically needs to consider at least $q$ interactions of different (polynomial) order, and tune the couplings to reach the right subspace of codimension $q$  in the coupling phase space. 
It is possible to interpret the interactions of higher order as certain hard-dimer configurations on the random lattices obtained from only the lowest-order interactions. In the case of matrix models and 2d gravity, coupling random planar diagrams to hard-dimers leads to the Yang-Lee edge singularity in the continuum limit, a non-unitary conformal matter coupled to gravity \cite{Staudacher:1990gj}.
This motivates the study of multi-critical behaviour in the context of enhanced random tensor models. 
However, the entropy exponent $\gamma$ is in general just a first indicator of what the continuum theory can be. For instance, an exponent $\gamma=-1/3$  does not necessarily mean that the continuum theory is the Yang-Lee singularity, as the same exponent is recovered for the Ising model on random lattices \cite{Kazakov:1986ei,Kazakov:1989gw}, leading to unitary conformal matter coupled to 2d gravity in the continuum. 

In the transition regime of proliferating baby-universes multi-criticality is a particular subtle issue.
In \cite{Korchemsky:1992uj} it is argued that, tuning a matrix model to order-$n$ multi-criticality and adding a strong-enough polymerization (non-local) term $\Tr(M^2)^2$, the $1/3$ exponent corresponding to proliferating baby-universes could be changed to a 
\be
\gamma= \frac1{n+2}
\ee
multi-critical exponent.
It is further argued in \cite{AlvarezGaume:1992np} that the exponent 
\be
\gamma= \frac m{n+m+1}
\ee
could be reached using non-local (multi-trace) interactions of higher order.
We will show that the latter is not realized in the tensor models considered here.

\para{The scalings $s_\cB$}

In order for the $1/N$ expansion of $2n$-point functions to be defined, the scalings $s_\cB$ in \eqref{eqref:PartFunct} have to be chosen such that the dependence  of the amplitude of Feynman diagrams in $N$ is bounded by $N^{f(n)}$, where $f(n)$ only depends on $n$. For the free-energy, we expect $f(n)=D$, $D$ being the rank of the tensor $T$, as $N^D$ is the number of degrees of freedom.
However, any choice leading to bounded amplitudes is not considered a good choice: if there are finitely many contributions to the leading order, the large $N$ correlation functions do not exhibit any singularity, and the theory is trivial in the continuum. 
Gurau proved \cite{Gurau:2014ch} that the choice $s_\cB=-\delta(\cB)\le0$, $\delta(\cB)$ being the Gurau degree of the bubble (see \cite{Gurau:2014ch}), leads to a defined $1/N$ expansion. The interaction bubbles with vanishing Gurau degree have an almost vector-like structure, and are called \emph{melonic} 
(\eg interactions labelled $2_\text{M},3_\text{M},3'_\text{M}$ in Table \ref{table:rank3-interactions},~\ref{table:rank4-interactions}). 
We do not describe them precisely, but for our purposes, the important properties we need are:
\begin{itemize}
\item[--] the only scaling leading to a defined and non-trivial $1/N$ expansion for a melonic interaction is $s_\text{melonic}=0$,
\item[--] the leading order of melonic random tensor models is tree-like, the 2-point function has a square-root singularity, so that the critical exponent is $\gamma=1/2$, and they lead to branched polymers in the continuum \cite{Gurau:2013th}.
\end{itemize}
A non-trivial consequence of a result from \cite{Gurau:2016gq} is that for every non-melonic case, 
the choice $s_\cB=-\delta(\cB)$ leads to a finite number of Feynman diagrams contributing to each order of the $1/N$ expansion, and in particular to a trivial leading order. 
For known cases of bubbles $\cB$ there is only one particular integer value of $s_\cB$ satisfying the two conditions above, and one has to look at each model individually to determine it. This value is positive for non-melonic interactions. Models with a choice of scaling leading to a defined and non-trivial $1/N$ expansion are known as {\it enhanced tensor models} in the literature \cite{Bonzom:2015gt, Bonzom:2016to}. 

\

The aim of this paper is to compute the appropriate scaling $s_\cB$ and the critical exponent $\gamma$ for every type of quartic and sextic interaction $\cB$ in dimension 3 and 4. The results we obtain are combined in Tables~\ref{table:rank3-interactions} and~\ref{table:rank4-interactions}. As will be clear in the following section, some interactions only differ by exchanging colours. In the tables, we only represent one interaction of each type, but we provide in the second column the number of interactions which only differ by an exchange of colours. In this way, the number of sextic interactions of rank 3 and 4 sums up to 7 and 41 respectively, in agreement with \cite{BenGeloun:2013ty}.

We recall that the amplitude of a vacuum diagram $\Gamma$ is given by (\ref{eqref:amp}), up to a symmetry factor. The symmetry factor is one for the two-point function. 
The accurate scaling $s_\cB$ leading to a defined and non-trivial $1/N$ expansion is deduced from the number of faces (\ref{eqref:Faces}) of leading-order diagrams, and the critical exponent $\gamma$ is given by the critical behaviour of the resulting leading-order two-point function, expected to be an indicator for the continuum limit as explained.

In this paper, we have chosen to use the stranded representation of interactions leading to diagrams which generalize ribbon graphs. Another graphical representation is commonly used in the tensor models literature, in which tensors $T$ are represented as white vertices, tensors $\bar T$ are represented as black vertices, and a line of colour $c$ is drawn between a black vertex and a white vertex if a strand of colour $c$ goes between the corresponding tensors $T$ and $\bar T$ in the stranded interaction. We represent these graphs as well in the third column of the tables. 
Finally, we represent the corresponding intermediate-field vertex which will be explained in detail in Section~\ref{sec:LOdiagrams}.

\subsection{Phase transition from a single type of interaction: quartic necklaces}
\label{sec:necklace}


The leading order of matrix models with single-trace interactions of orders $\{k_1, k_2, \cdots\}$ involves planar ribbon graphs with vertices that have $\{k_1, k_2, \cdots\}$ incident propagators. 
The critical behaviour of the correlation functions is given by the exponent $\gamma=-1/2$, or, if the couplings are fine-tuned, by the multi-critical exponents $-1/m$. It is still possible to obtain other critical regimes, such as that of proliferating universes, characterized by the exponent $\gamma=1/3$, but one needs to consider multi-trace interactions \cite{Das:1990gp,AlvarezGaume:1992np}, graphically represented by non-connected stranded interactions. Here, we introduce a one-tensor model with \emph{connected} stranded interactions, which presents a branched-polymer phase, a 2d planar phase, and a proliferation of baby universes at the phase transition between them. It happens to be equivalent to a multi-matrix $N^2\times N^2$ hermitian model (with usual trace). This tensor model has three interactions, that previously introduced in (\ref{fig:Neck}) and two obtained by exchanging colours.
\be 
\label{fig:TwoNecklaces}
\bnp2^{12} \ \ 
\begin{array}{c}
\includegraphics[scale=0.6]{diagrams/Neck.pdf}\end{array}
\hspace{2cm}
\bnp2^{13} \ \ 
\begin{array}{c}\includegraphics[scale=0.6]{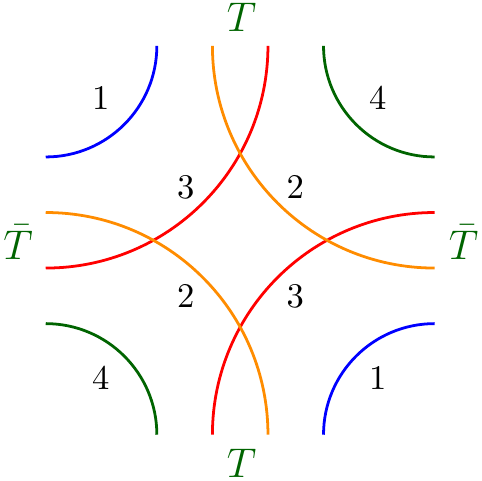}\end{array}
\hspace{2cm}
\bnp2^{14} \ \ 
\begin{array}{c}\includegraphics[scale=0.6]{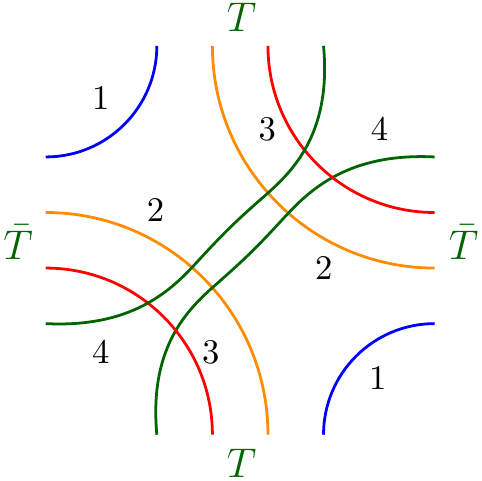}\end{array}
\ee 
These interactions have been called (quartic) \emph{necklaces} in \cite{Bonzom:2015gt} because of their representation in graphs (see the second and fifth rows of Table~\ref{table:rank4-interactions}). We conserve this denomination here. The corresponding partition function is
\be 
\cZ_{4}(\tn2,\tn3,\tn4, N)=\int_{\bC^D}\e^{\tn2 N^4  \Tr_{\bnp2^{12}}(T,  \bar{T}) + \tn3 N^4  \Tr_{\bnp2^{13}}(T,  \bar{T})+\tn4 N^4  \Tr_{\bnp2^{14}}(T,  \bar{T})}d\mu_{0}(T,  \bar{T}),
\ee
with coupling $\tn i=\tnp2^{1i}$ corresponding to the interactions $\bnp2^{1i}$. We know from \cite{Bonzom:2015gt} that the appropriate scaling is $\sb=1$. The amplitude (\ref{eqref:amp}) rewrites as
\be
\cA(\gG)\sim \tn2 ^{\B_2(\gG)} \tn3 ^{\B_3(\gG)} \tn4 ^{\B_4(\gG)} N^{F(\gG)-2\B(\gG)},
\ee
where $\B_i$ denotes the number of interaction bubbles of type $\bnp2^{1i}$, and $\B=\B_2+\B_3+\B_4$.

A simple argument shows that the random tensor model for only a single non-vanishing coupling,
\eg $\tn3=0$, $\tn4=0$, 
is described by a one-matrix model: by grouping together the colours 1,2, and 3,4 in (\ref{fig:Neck}), one obtains ribbon graphs for a matrix $M_{(12)(34)}=T_{1234}$ in which strands carry two colours instead of one. 
\be 
\begin{array}{c}\includegraphics[scale=0.6]{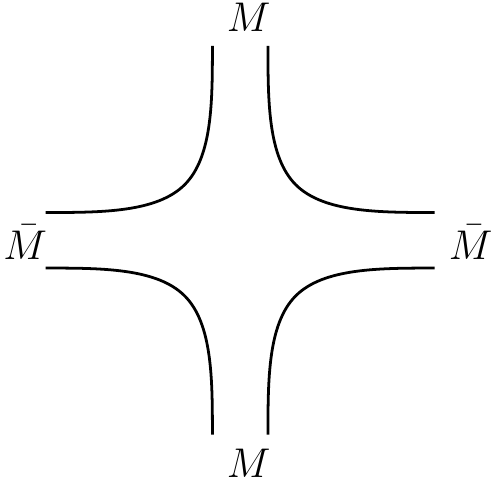}\end{array}
\hspace{3cm}
\begin{array}{c}\includegraphics[scale=0.6]{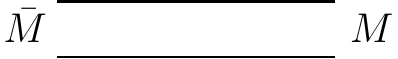}\end{array}
\ee 
Such ribbon graphs label the perturbative expansion of a quartic $N^2\times N^2$ matrix model, with partition function
\be 
\cZ_{4}(\tn2, N)=\int \frac{dMd\bar M}{\pi^{N^2}} \e^{-\Tr MM^\dagger - \tn2 N^{-2} \Tr MM^\dagger M M^\dagger}.
\ee

In the case where not all three couplings vanish, the tensor model can also be rewritten as a matrix model. Indeed, it was shown in \cite{Bonzom:2015gt} that the Feynman diagrams of this theory are one-to-one with embedded diagrams, or ribbon graphs, with vertices of any valency (the valency is the number of incident propagators), and three kinds of propagators, those carrying colours 1 and 2 and coming with a factor $\tn2$, those carrying colours 1 and 3 and coming with a factor $\tn3$, and those carrying colours 1 and 4 and comming with a factor $\tn4$. Embedded diagrams label the perturbative expansion of matrix models, and the model turns out to be described by the following three-matrix model
\be 
\cZ_{4}(\tn2,\tn3,\tn4, N)=\int \e^{- \Tr\ln\bigl(\un^{\otimes 4} - i\sqrt {2\tn2} M_{12}\otimes\un_{34} - i\sqrt {2\tn3} M_{13}\otimes\un_{24}- i\sqrt {2\tn4} M_{14}\otimes\un_{23}\bigl)}d\nu(M_{12},M_{13}, M_{14}),
\ee
where $M_{12}$, $M_{13}$, and $M_{14}$ are $N^2\times N^2$ hermitian matrices that propagate their subscript colours, $\un_{ij}$ is the identity in colours $i$ and $j$, and the measure is 
$d\nu(M_{12},M_{13}, M_{14})={dM_{12}dM_{13}dM_{14}}/{\cZ_0} \, e^{-\frac{N^2}2\Tr( M_{12}^2+M_{13}^2+M_{14}^2) } $, $\cZ_0$ being chosen so that $\cZ_{4}(0,0,0, N)=1$.
This is the intermediate-field representation of the tensor model. 
Denoting $E_{1i}$ the number of propagators of type $1i$, the amplitude of a ribbon graph $M$ is 
\be
\cA(M)\sim \tn2 ^{E_{12}(M)}\tn3 ^{E_{13}(M)}\tn4 ^{E_{14}(M)}N^{F(M)-2E(M)},
\ee
where now $F(M)=\sum_{i=1}^4 F_i(M)$ is the sum of the faces of colours 1, 2, 3 and 4 of the embedded diagram, and $E$ is the total number of propagators.
Leading-order diagrams are planar, but also satisfy an additional constraint on the possible loops, which depends on the colours of the propagators.

The usual three regimes 
also appear in this case, as we show in Subsection~\ref{subsec:TwoNecks}: for two vanishing couplings, 
this reduces to a usual matrix model, in the universality class of pure 2d gravity ($\gamma=-1/2$); for  $\tn2=\tn3=\tn4$, there is a destructive interference between the three kind of matrices which reduces the critical behaviour to a square root singularity, with exponent $\gamma=1/2$. Some intermediary values of the coupling constant reproduce the regime of proliferating baby universes, with exponent $\gamma=1/3$. 

The only other type of quartic interactions in dimension 4 is melonic (see Table~\ref{table:rank4-interactions}). The quartic melonic model is the most studied uncoloured tensor model.
As all melonic models, it has a tree-like leading order characterized by the critical exponent $\gamma=1/2$ and converging towards branched polymers in the continuum limit. Coupling a quartic melonic interaction and an interaction of the kind in \eqref{fig:TwoNecklaces}, the same three regimes appear, as detailed in \cite{Bonzom:2015gt}.

\subsection{The enhanced $K_{3,3}$ tensor model}

Before considering all sextic interactions for rank $\rk=4$ tensors, it is useful to understand the sextic interactions at rank $\rk=3$. There are 3 different kind of sextic interactions in dimension 3. Two of them are melonic (type $3_\textrm{M}$ and $3'_\textrm{M}$ interactions in Table~\ref{table:rank3-interactions}), and therefore are not to be enhanced ($s=0$). Their critical behaviour is known to be that of branched polymers. 
We therefore focus on the 3 dimensional theory with the following non-melonic interaction
\be 
\includegraphics[scale=1]{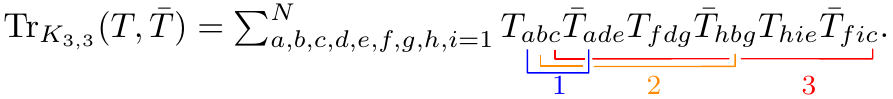}
\ee 
and corresponding partition function 
\be 
Z_{\ktt}(\tk, N)=\int_{\bC^D}e^{\tk N^3  \Tr_{\ktt}(T,  \bar{T})}d\mu_{0}(T,  \bar{T}).
\ee
It is represented below, as well as the corresponding propagator. 
\be 
\label{fig:K33}
\begin{array}{c}\includegraphics[scale=0.6]{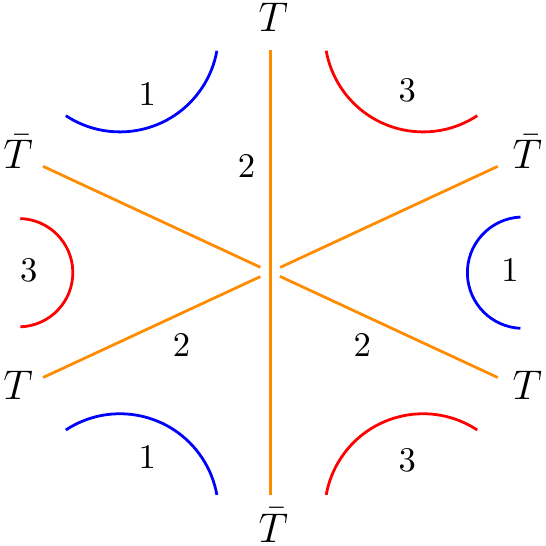}\end{array}
\hspace{3cm}
\begin{array}{c}\includegraphics[scale=0.6]{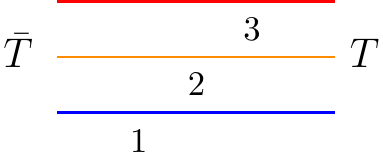}\end{array}
\ee 
It is referred to as the $K_{3,3}$ interaction in the random tensor literature, as one obtains the complete bipartite $K_{3,3}$ graph by replacing each tensor by a vertex. It is a natural model to consider as it is symmetric upon exchange of colours and no non-planar interactions have been considered before (you can not draw it without a crossing). It has been shown in  \cite{SWMaps} that the number of faces \eqref{eqref:Faces} of leading-order diagrams is 
\be
F_\text{LO} = 3(1+\B),
\ee
where $\B$ is the number of interaction bubbles. From \eqref{eqref:amp}, this implies that the strongest scaling one can choose is $s=1$. 
Knowing this, the amplitude of Feynman diagrams  (\ref{eqref:amp}) rewrites as
\be
\cA(\gG)\sim \tk ^{\B(\gG)}N^{F(\gG)-3\B(\gG)}.
\ee
In the large $N$ limit, only diagrams which maximize $F(\gG)-3V(\gG)$ survive. They have been identified in \cite{SWMaps}, however, although a $\gamma=1/2$ exponent was expected, it had not been computed before. They scale in $\cO(N^3)$ and we show in Section~\ref{subsec:K33LO} that the leading-order two-point function satisfies the following self-consistency equation,
\be 
\gt(\tk)=1+3\tk \gt\bigl(\tk)^3+ 3\tk^2\gt(\tk)^6.
\ee 
Using standard results in singularity analysis of tree generating functions \cite[ch. VII.4]{Flajolet:2009wm} we find that the solution to this equation has a square-root expansion
\be 
\gt(\tk) = \gt(\tc) - c \sqrt{1-\tk/\tc}+o\bigl(\sqrt{1-\tk/\tc}\bigr),
\ee
indeed leading to a $\gamma=1/2$ exponent, and where $\tc>0$, $\gt(\tc)>0$, and $c>0$ can be computed numerically.
%
 To summarize, we present all interactions of order 6 for rank $\rk = 3$ in table \ref{table:rank3-interactions}.

\begin{table}
\begin{tabular}{|M{1.5cm}|M{1.5cm}|M{3cm}|M{3cm}|M{3cm}|M{1.5cm}|M{1.5cm}|} 
\hline
%
{\bf Name} & {\bf Numb. of col.} & {\bf Interaction} & {\bf Coloured graph}  & {\bf Int Field} & $\quad s_B\quad$ & $\ \gamma_B\ $  \\
\hline
$2_\textrm{M}$ & 3 &
\includegraphics[scale=0.4]{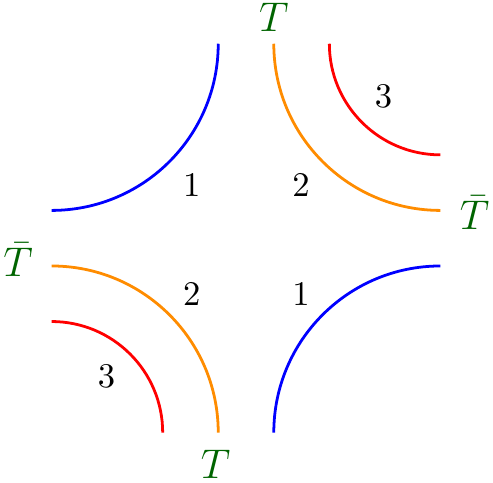} &
\includegraphics[scale=0.5]{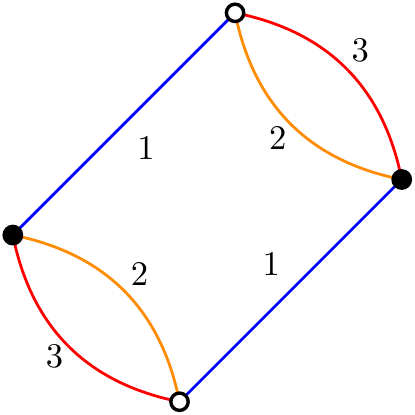} &
\includegraphics[scale=0.5]{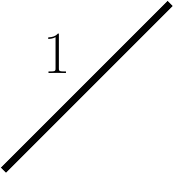}
& 0 &  $ \frac 1 2 $\\
\hline
$3_\textrm{M}$ & 3 
& \includegraphics[scale=0.4]{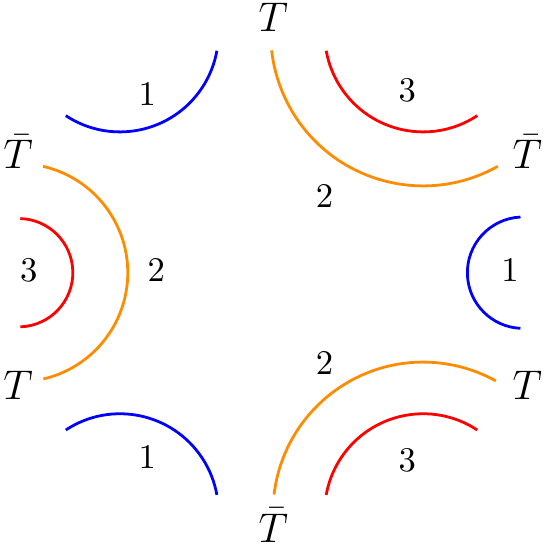}
& \includegraphics[scale=0.5]{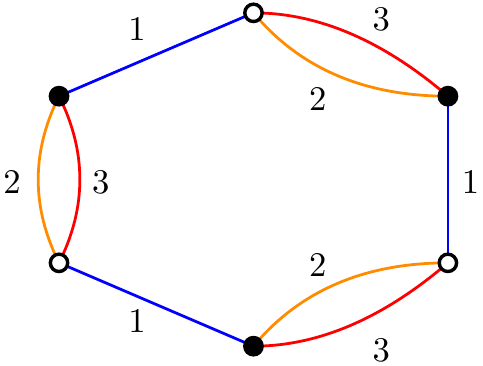}
& \includegraphics[scale=0.5]{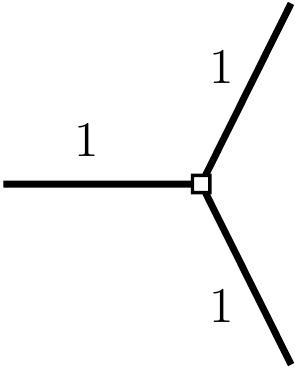} & 0 &   $ \frac 1 2 $\\
\hline
$3'_\textrm{M}$ & 3 
&\includegraphics[scale=0.4]{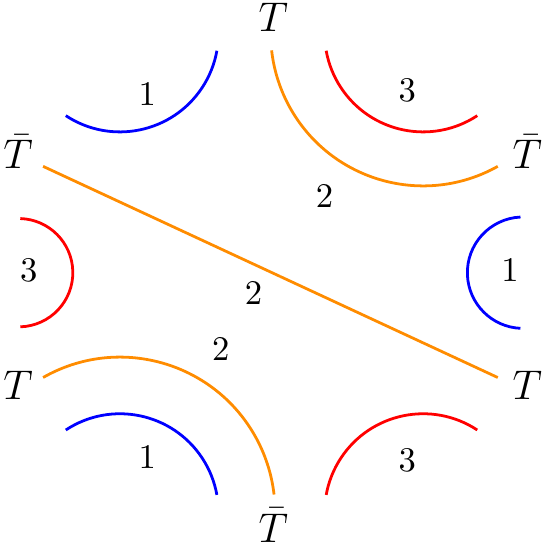} 
& \includegraphics[scale=0.5]{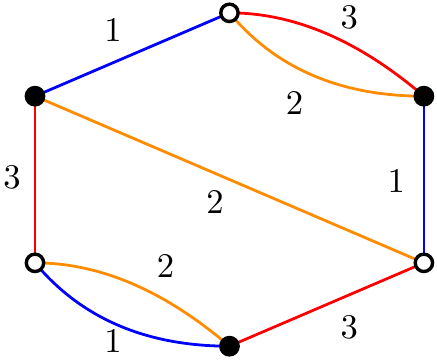} & \includegraphics[scale=0.5]{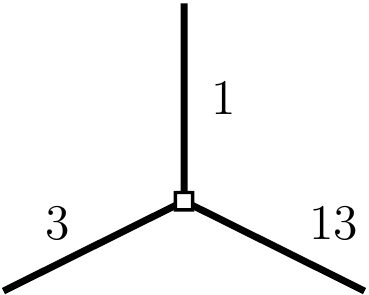} & 0 &    $ \frac 1 2 $\\
\hline
$3_\textrm{K}$ & 1 &\includegraphics[scale=0.4]{diagrams/K33.pdf} & \includegraphics[scale=0.5]{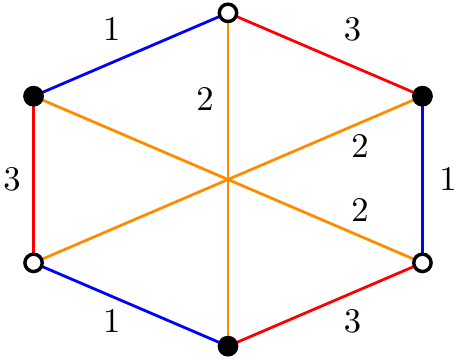} & \includegraphics[scale=0.5]{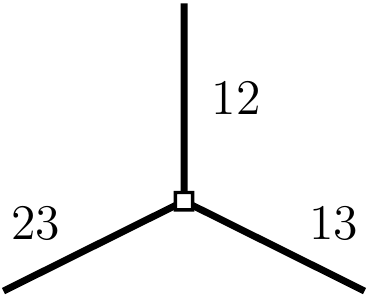} & 1 &   $ \frac 1 2 $\\
\hline
\end{tabular}
\caption{\label{table:rank3-interactions}
All $U(N)^3$-invariant interactions of order 4 and 6 in rank-3 tensors. The non-local structure of the interaction is shown as stranded diagram, in the coloured-graph representations common in the tensor-model literature and in the intermediate-field representation explained below in section \ref{sec:LOdiagrams}.}
\end{table}

\subsection{Order six theory in four dimensions}
\label{subsec:Ord6Rg4Intro}

In dimension 4, besides melonic interactions $3_\textrm{M}$ and $3'_\textrm{M}$ (see Table~\ref{table:rank4-interactions}), there are necklace interactions $3_\textrm{N}$, and the interactions of type $3_\textrm{MN}$, which all generate leading orders similar to that of quartic necklaces, there is the interactions $3'_\textrm{N}$, which, as we will explain in Subsection~\ref{subsection:K334}, can be understood as an enrichment of the quartic necklaces, and  there are two kinds of interactions, $3_\textrm{K}$ and $3'_\textrm{K}$, which are obtained from the $\ktt$ interaction by adding an additional colour. As detailed in Subsection~\ref{subsection:K334}, it is very easy to show that interaction of the type $3'_\textrm{K}$ have a tree-like leading order with exponent $\gamma=1/2$. We therefore describe the $3_\textrm{K}$ tensor model here.
The interaction is obtained from the $K_{3,3}$ interaction by adding strands of colour $4$ parallel to the strands of colour $i$, where $i=1,2,$ or $3$ (and by possibly exchanging colours). An interaction of this type is
\be 
\begin{array}{c}\includegraphics[scale=0.4]{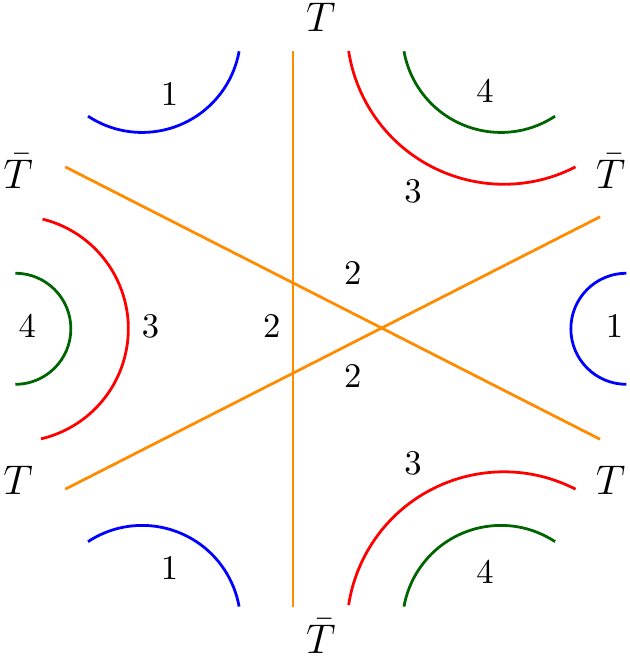}\end{array} \hspace{1.5cm}
\mathrm{Tr}_{\cB_{3_\textrm{K}}}(T,\bar T)=\sum_{a,b,c,d,e,f,g,h,i, j, k, l=1}^N T_{abcd}\bar T_{aefg}T_{heij}\bar T_{kbij}T_{klfg}\bar T_ {hlcd}.
\ee 
 We will study the critical regimes for an order six theory that also includes the following type $3_\textrm{M}$ melonic interaction,
\be 
\label{eqref:Melo6}
\begin{array}{c}\includegraphics[scale=0.4]{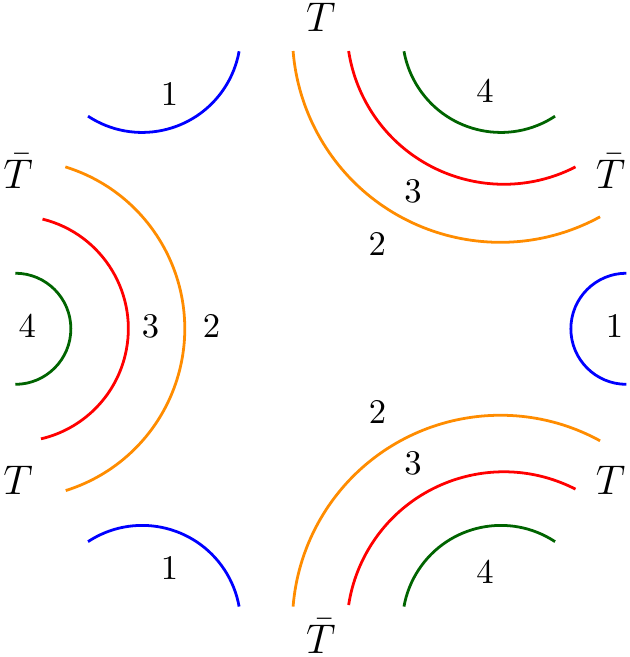}\end{array}\hspace{1.5cm}
\mathrm{Tr}_{\bmp3}(T,\bar T)=\sum_{a,b,c,d,e,f,g,h,i, j, k, l=1}^N T_{abcd}\bar T_{aefg}T_{hefg}\bar T_{hijk}T_{lijk}\bar T_ {lbcd}.
\ee 

The corresponding partition function is
\be 
Z_{6}(\lambda, N)=\int_{\bC^D}e^{\tk N^{3+s_{3_\textrm{K}}}  \Tr_{3_\textrm{K}}(T,  \bar{T}) + \tmp3 N^3  \Tr_{\bmp3}(T,  \bar{T})}d\mu_{0}(T,  \bar{T}).
\ee
Because the interaction in \eqref{eqref:Melo6} is melonic, we should have a square-root critical behaviour in $\tmp3$, when $\tk=0$. We will show in Subsection~\ref{subsection:K334} that the right scaling is $s_{3_\textrm{K}}=2$. We will also show that the three critical regimes with exponents $1/2$, $-1/2$, and $1/3$ are obtained for this model.

\begin{table}
\begin{tabular}{|M{1.5cm}|M{1.5cm}|M{3cm}|M{3cm}|M{3cm}|M{1.5cm}|M{1.5cm}|} 
\hline
%
{\bf Name} & {\bf Numb. of col.} & {\bf Interaction} & {\bf Coloured Graph}  & {\bf Int Field} & $\quad s_B\quad$  & $\ \gamma_B\ $  \\
\hline
$2_\textrm{M}$ & 4 & \includegraphics[scale=0.4]{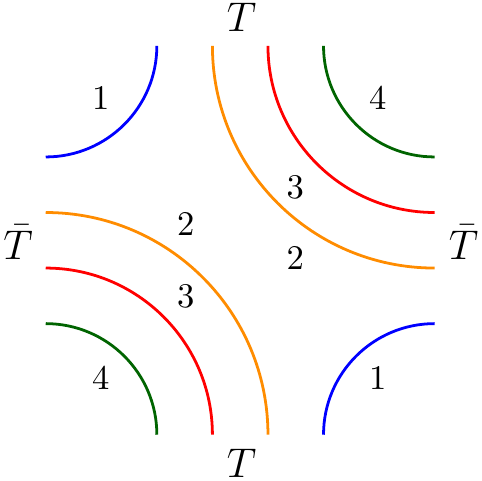} & \includegraphics[scale=0.5]{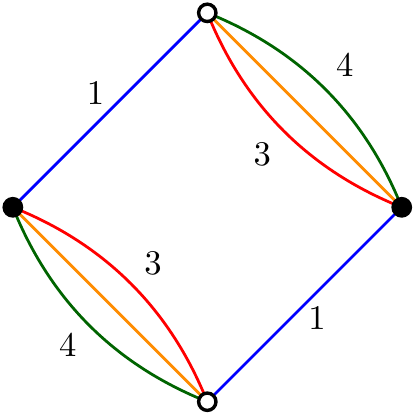} & \includegraphics[scale=0.5]{diagrams/Melo4IF.pdf}   & 0 &   $ \frac 1 2 $\\
\hline
$2_\textrm{N}$ & 3 & \includegraphics[scale=0.4]{diagrams/Neck.pdf} & \includegraphics[scale=0.5]{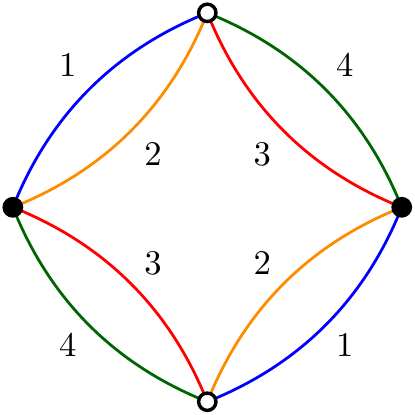} &
\includegraphics[scale=0.5]{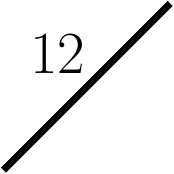} & 1 &   {\begin{tabular}{@{}c@{}}  $  -1/ 2$\\[+2ex]$  1/ 2$\\[+2ex] $1/3$ \end{tabular}}\\
\hline
$3_\textrm{M}$ & 4 & \includegraphics[scale=0.35]{diagrams/Melo6S.pdf} & \includegraphics[scale=0.5]{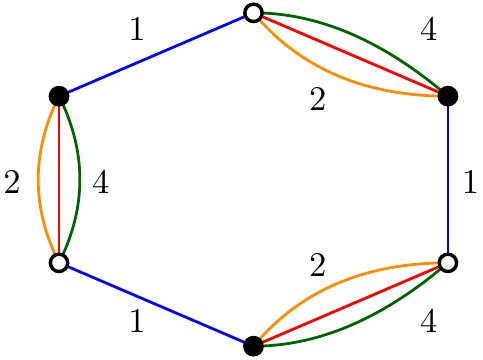} & \includegraphics[scale=0.5]{diagrams/Melo26IF.pdf} & 0 &  $ \frac 1 2 $\\
\hline
$3_\textrm{M}'$ & 6 &\includegraphics[scale=0.35]{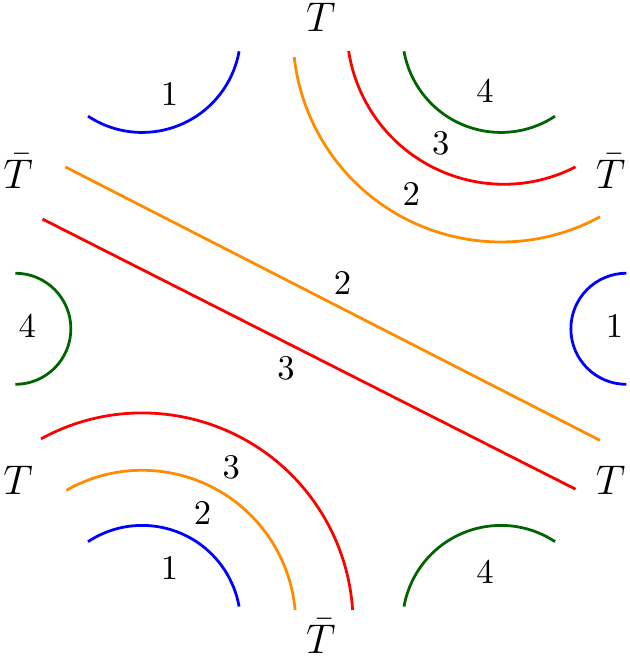} & \includegraphics[scale=0.4]{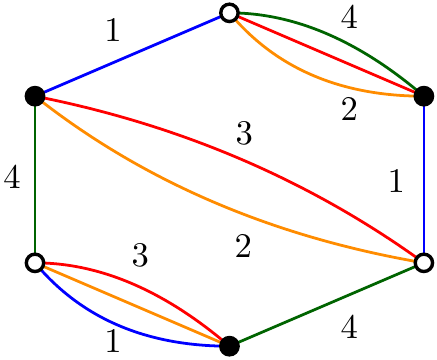} & \includegraphics[scale=0.5]{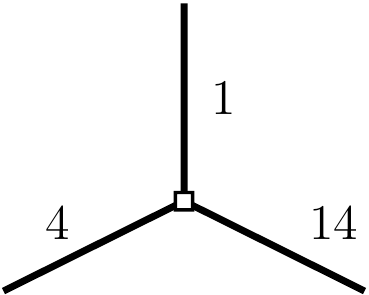}  & 0 &    $ \frac 1 2 $\\
\hline
$3_\textrm{N}$ & 3 &\includegraphics[scale=0.35]{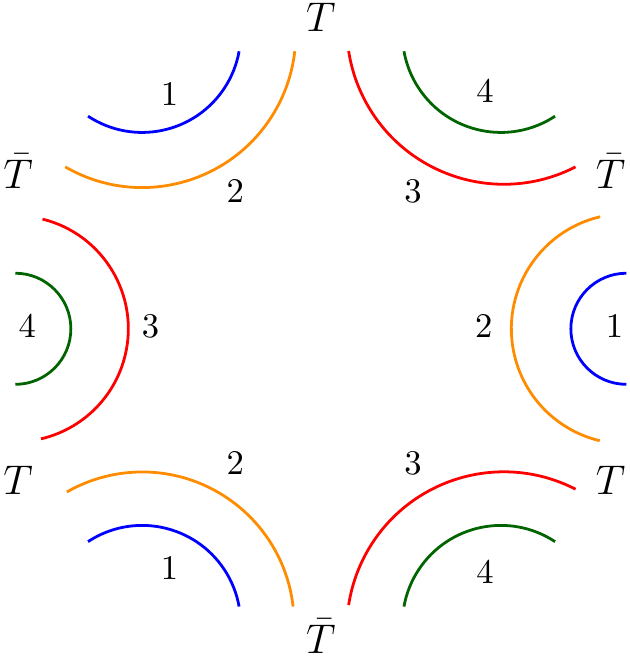} & \includegraphics[scale=0.5]{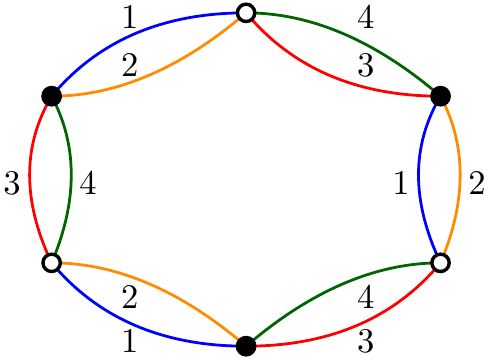} & \includegraphics[scale=0.5]{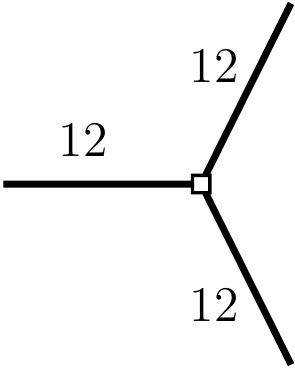} & 2 &   {\begin{tabular}{@{}c@{}}  $  -1/ 2$\\[+2ex]$  1/ 2$\\[+2ex] $1/3$ \end{tabular}}\\
\hline
$3'_\textrm{N}$ & 6 &\includegraphics[scale=0.35]{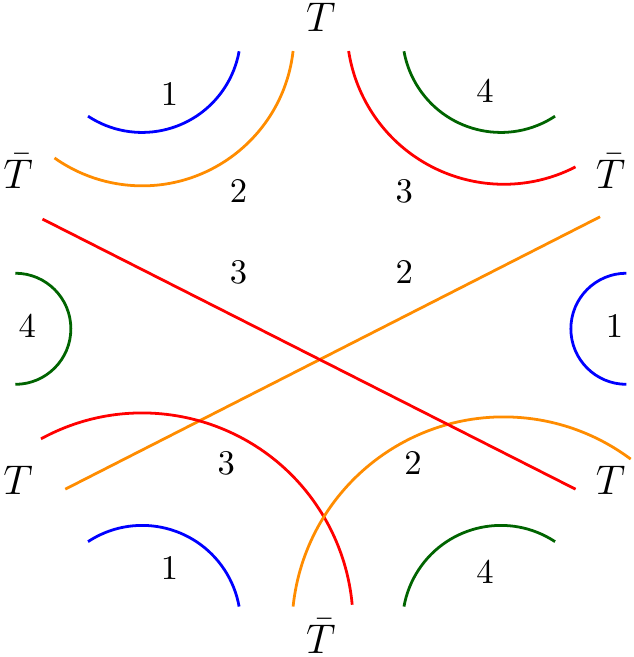} & \includegraphics[scale=0.5]{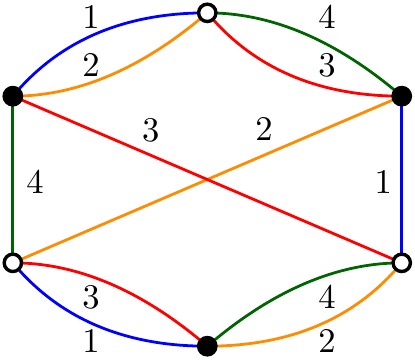} & \includegraphics[scale=0.5]{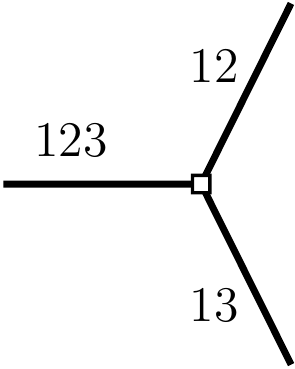} & 2 &   $ \frac 1 2 ?$\\
\hline
$3_\textrm{MN}$ & 12 
& \includegraphics[scale=0.35]{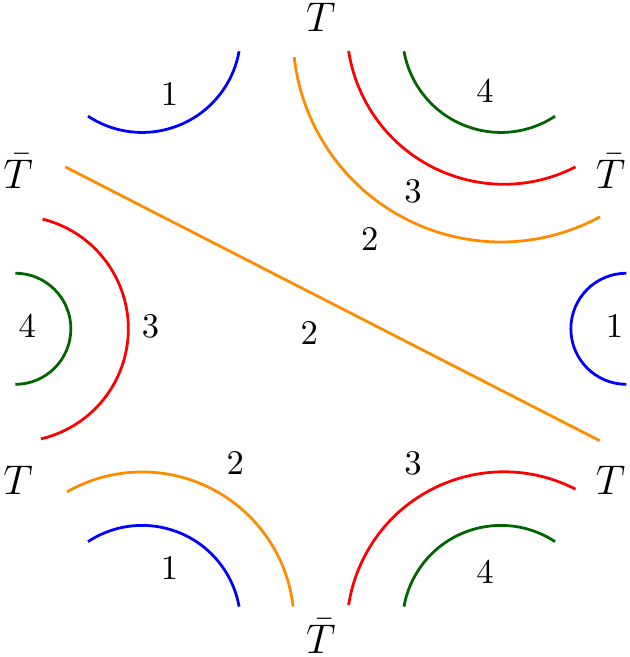} 
& \includegraphics[scale=0.5]{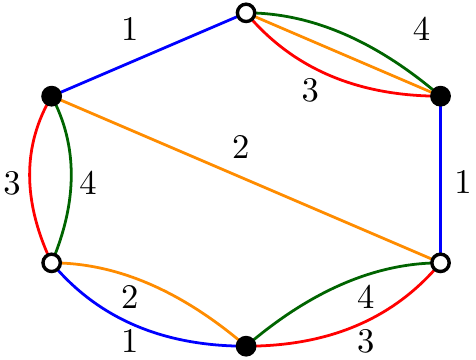} 
& \includegraphics[scale=0.5]{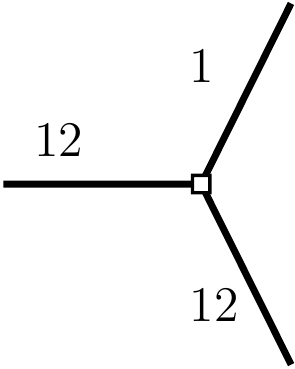} & 1  
&  {\begin{tabular}{@{}c@{}}  $  -1/ 2$\\[+2ex]$  1/ 2$\\[+2ex] $1/3$ \end{tabular}}\\
\hline
$3_\textrm{K}$ & 6 &\includegraphics[scale=0.35]{diagrams/K334S.pdf} & \includegraphics[scale=0.5]{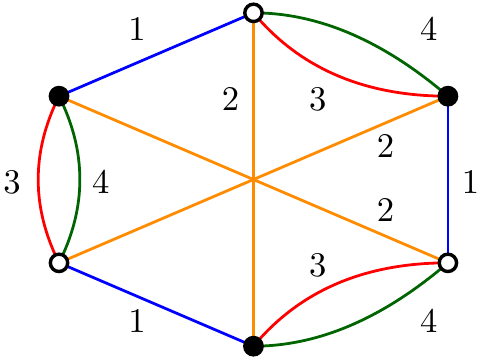} & \includegraphics[scale=0.5]{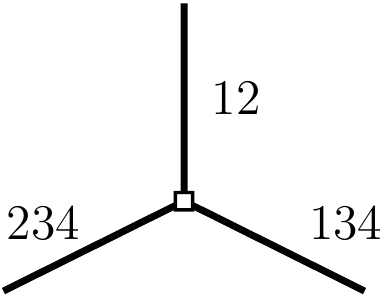}   & 2 &    $ {\begin{tabular}{@{}c@{}}  $  -1/ 2$\\[+2ex]$  1/ 2$\\[+2ex] $1/3$ \end{tabular}} $\\
\hline
$3'_\textrm{K}$ & 4 &\includegraphics[scale=0.35]{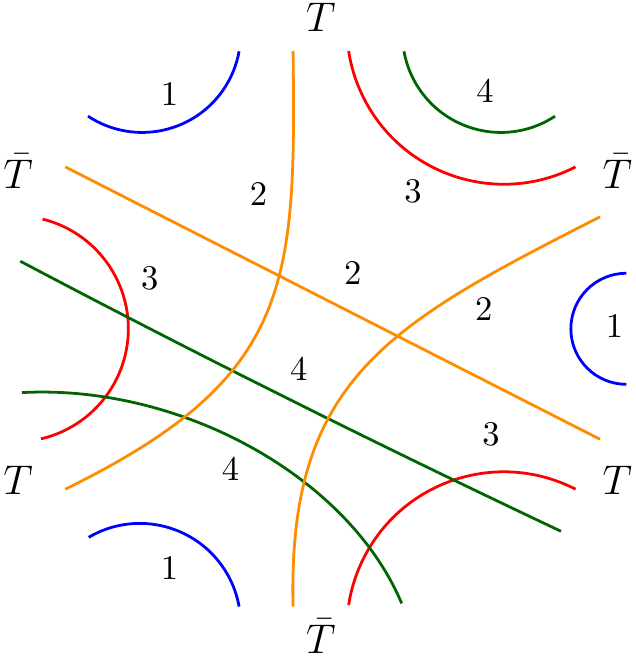} & \includegraphics[scale=0.5]{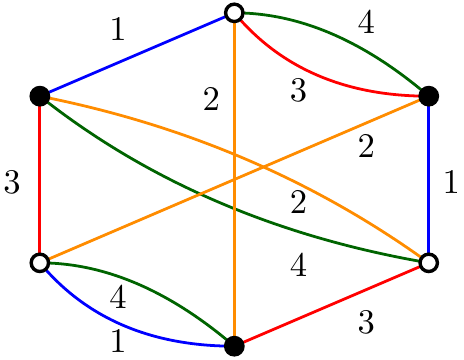} & \includegraphics[scale=0.5]{diagrams/K33TreeIF.pdf}  & 1 &    $ \frac 1 2 $\\
\hline
\end{tabular}
\caption{\label{table:rank4-interactions}
All $U(N)^4$-invariant interactions of order 4 and 6 in rank-4 tensors, shown as stranded diagram, coloured-graph 
and in the intermediate-field representation explained below in section \ref{sec:LOdiagrams}.}
\end{table}

\newpage


\section{Leading-order diagrams}
\label{sec:LOdiagrams}

In this section we explain the derivation of the Dyson-Schwinger equations for the tensor models under consideration by counting their Feynman diagrams.
The evaluation of the perturbative sum of the 2-point function can be performed most efficiently transforming the tensor-model state sum to an \emph{intermediate-field representation} involving embedded diagrams \cite{Delepouve:2014wk, Gurau:2014kj, SWMaps}. This was illustrated in section \ref{sec:necklace} for the quartic tensor model with a single interaction of type  $\bnp2^{1i}$, for which the Feynman diagrams are in one-to-one correspondence with four-valent ribbon graphs, 
and in which closed strands carry a weight $N^2$. For other models, there generally is no one-to-one map as simple but one can develop more subtle transformations. 

There is a one-to-one correspondence between Feynman diagrams and embedded diagrams equivalent to ribbon graphs, introduced in \cite{SWMaps}, in which interactions are replaced by white vertices of a specific kind.  
White vertices can only be linked to black vertices, and conversely. 
Propagators, which we will call \emph{edges} in this section, carry various colour-sets $\cI\subset\{1,\cdots,D\}$ that depend on the interactions, and black vertices have no restriction on possible incident edges. 
Importantly, embedded diagrams, also sometimes called fatgraphs, are graphs for which the cyclic ordering of edges around vertices matters: two diagrams which only differ by the ordering of edges around vertices are different diagrams. Consequently, two half-edges incident to the same vertex and following each other are separated by a well-defined corner. 
A face in an embedded diagrams is a closed walk: a cyclic succession of corners that are on the same side of the same edge (they correspond to the closed strands of the equivalent ribbon graph).  The propagators in the stranded representation are mapped to corners in the new intermediate-field representation, and crucially, the faces of colour $i$ of the stranded diagrams (\eqref{eqref:Faces}) are mapped to faces around the embedded diagrams obtained by keeping all the black vertices, and only the edges whose colour-set contains colour $i$, as can be seen on the example of (\ref{fig:ExBij}). We can therefore apply ribbon graphs tools to solve tensor models. We refer the reader to the Appendix for a more detailed explaination of this correspondence.

To obtain the white vertex corresponding to a given interaction, one has to choose a way to pair every tensor $T$ with a $\bar T$. The intermediate-field vertices corresponding to each interaction are given in the fifth column of  Tables~\ref{table:rank3-interactions} and~\ref{table:rank4-interactions}. The choices of pairings
on order-6 interactions
made to obtain these embedded vertices are given by the following rules,
\be 
\label{fig:Pairings}
\begin{array}{c}\includegraphics[scale=0.45]{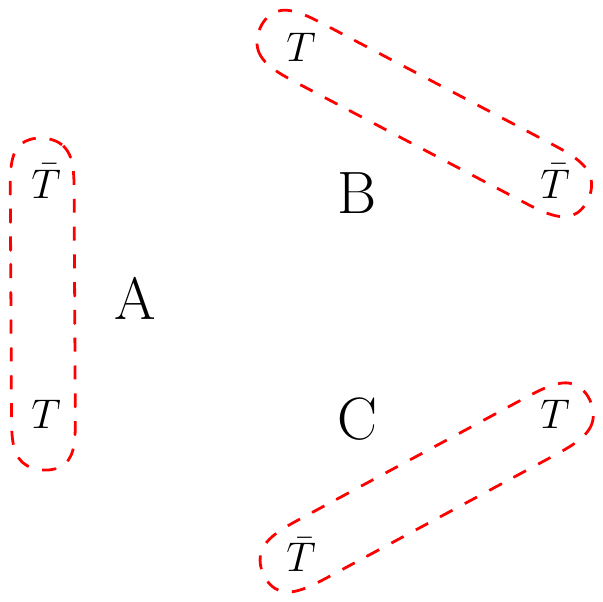}\end{array} \qquad\leftrightarrow\qquad
\begin{array}{c}\includegraphics[scale=0.5]{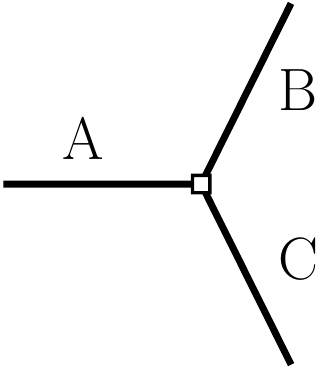}\end{array},
\qquad\text{and}\qquad
\begin{array}{c}\includegraphics[scale=0.45]{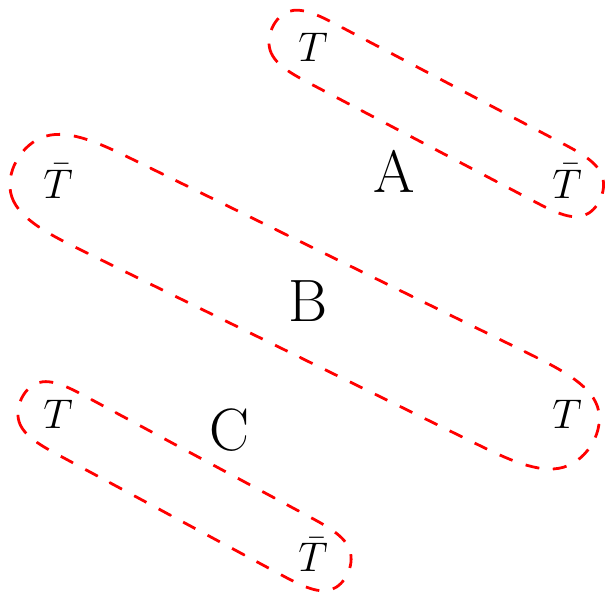}\end{array} \qquad\leftrightarrow\qquad
\begin{array}{c}\includegraphics[scale=0.5]{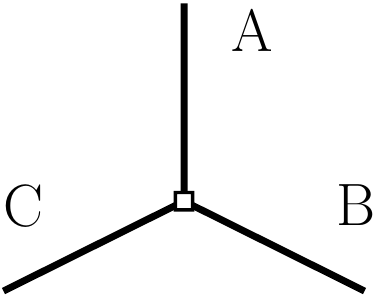}\end{array}.
\ee 
An example of a stranded diagram and the corresponding intermediate-field embedded diagram are shown below, \be
\label{fig:ExBij}
\begin{array}{c}\includegraphics[scale=0.3]{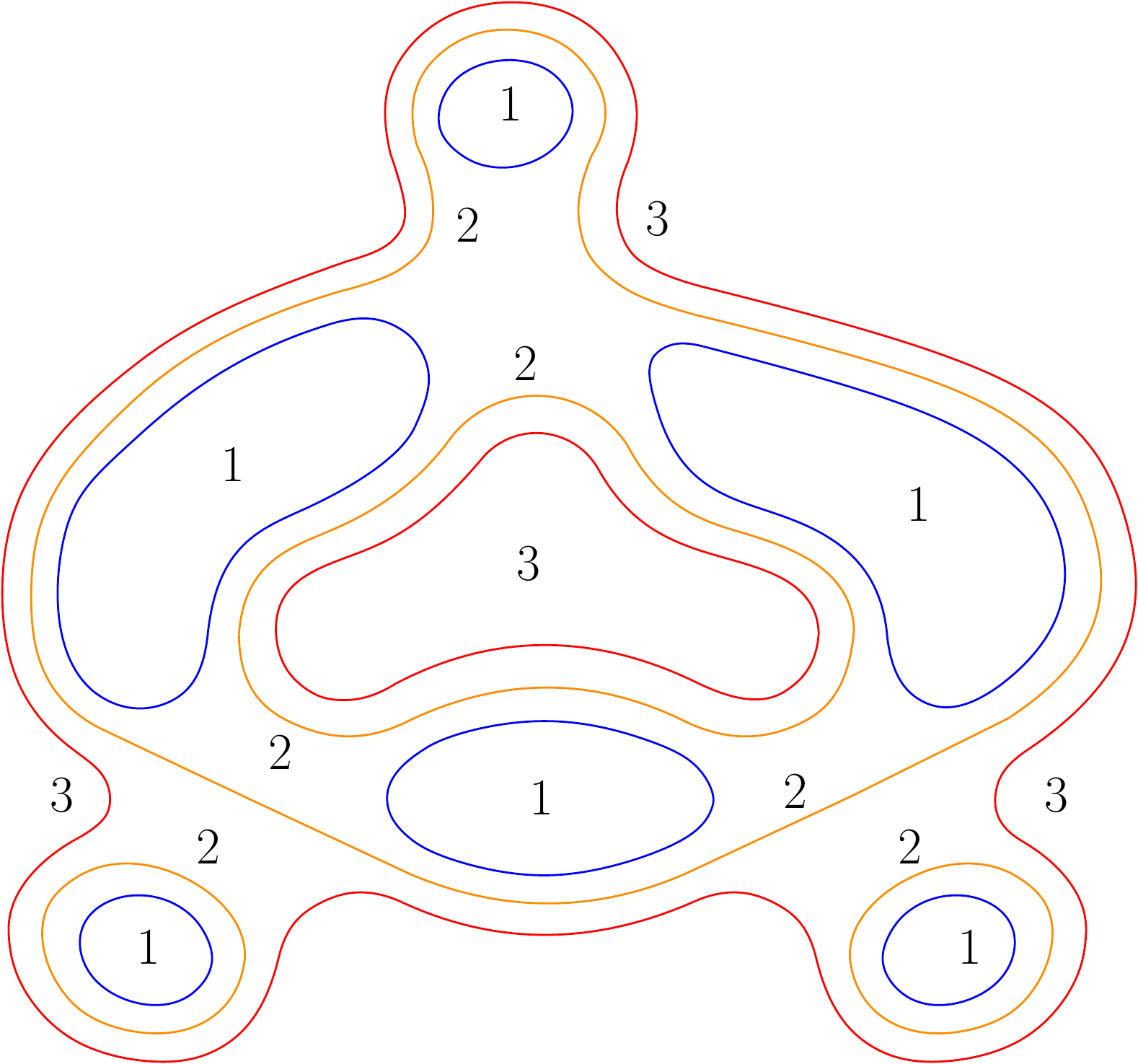}\end{array}
\hspace{0.9cm}\leftrightarrow\hspace{1.1cm}
\begin{array}{c}\includegraphics[scale=0.6]{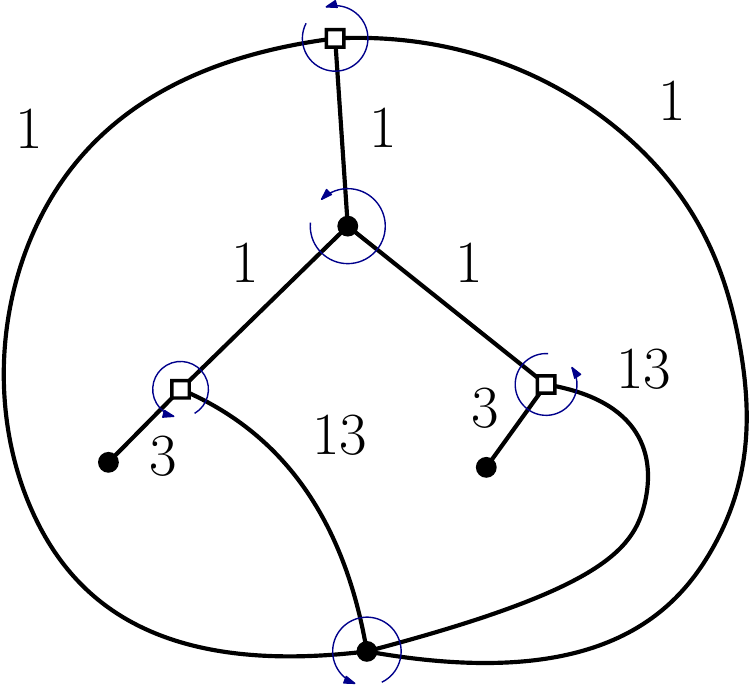}\end{array}.
\ee 
Given an embedded diagram $\cM$, we denote $L(\cM)$ its number of independent loops, given in terms of its number of edges $E$, vertices $V$, and connected components $K$ by 
\be
L(\cM) = E(\cM) - V(\cM) + K(\cM).
\ee
A tree is a diagram with no loops. The genus $g$ of an embedded diagram is given in terms of its number of faces $F$ by 
\be
2K(\cM) - 2g(\cM) = V(\cM) - E(\cM) + F(\cM).
\ee
It is the minimal number of holes of a surfaces on which the diagram can be drawn without crossings. If $\cM$ is an intermediate-field diagram, we may consider the colour-$i$ sub-diagram obtained by keeping all the black vertices and all the edges whose colour-set contains colour $i$. We respectively denote $L_i(\cM)$,  $F_i(\cM)$, and $g_i(\cM)$, the number of loops, faces, and the genus of the corresponding colour-$i$ sub-diagram. If $\cM$ is an intermediate-field diagram and $\cT$ is an intermediate-field tree with the same number $\B_\cB(\cM)$ of interaction bubbles of each type, then the following identity holds, which was shown in \cite{SWMaps},
\be
\label{eqref:TvsM}
F(\cT)-F(\cM) = DL(\cM) + 2\sum_{i=1}^D \bigl( g_i(\cM) - L_i(\cM) \bigr).
\ee
We will mainly use this formula to determine leading-order diagrams of the tensor models. The number of faces of trees is always linear in the number of interaction bubbles of each kind,
\be
\label{eqref:FTrees}
F(\cT) = D + \sum_{\cB\in\bB} a_\cB \B_\cB(\cT),
\ee
where $a_\cB$ only depends on the interaction bubble and the way the tensors $T$ and $\bar T$ have been paired (it is the number of faces when putting the propagators between the pairs $(T, \bar T)$, minus the rank of the tensor) \cite{SWMaps}. If trees belong to the leading order, then $a_\cB$ gives the strongest scaling $s_\cB$ one can choose to have a defined $1/N$ expansion 
\be
\label{eqref:AfromS}
s_\cB = (n_\cB -1)(D-1) - a_\cB.
\ee
All interactions of order 4 and 6 in dimensions 3 and 4 are such that, for the choices of intermediate-field vertices presented in Tables~\ref{table:rank3-interactions} and~\ref{table:rank4-interactions}, trees belong to the leading order, and we can therefore deduce the appropriate scalings from the number of faces of trees. With this choice of scaling, the amplitudes of intermediate-field diagrams scale as 
\be
\cA(\cM) \sim N^{D\bigl(1-L(\cM)\bigr) + 2\sum_{i=1}^D \bigl( L_i(\cM) - g_i(\cM) \bigr)} = \cO(N^D).
\ee

\subsection{Intermediate-field representation of rank $\rk=4$ quartic tensor models} 
\label{subsec:Quart}

For quartic theories with interactions of the type $2_\textrm{N}$ (\ref{fig:TwoNecklaces}) or of the type  $2_\textrm{M}$ (first row of Table~\ref{table:rank4-interactions}), the transformation can be done directly on the correlation functions using the Hubbard-Stratonovich transformation \cite{Stratonovich:1957, Hubbard:1959}. 
This can be seen e.g. in \cite{Delepouve:2017}.  However we here focus on describing this correspondence at the diagram level. We choose to pair a tensor $T$ and a $\bar T$, and we see the quartic interactions as new kind of edges between these two pairs\footnote{In the quartic case, the intermediate-field vertex would have two incident edges, and it can just be replaced with an edge.}, as illustrated in (\ref{fig:Bij}). The fact that interactions become edges is why this new representation is called the intermediate-field representation. \be 
\label{fig:Bij}
\begin{array}{c}\includegraphics[scale=0.5]{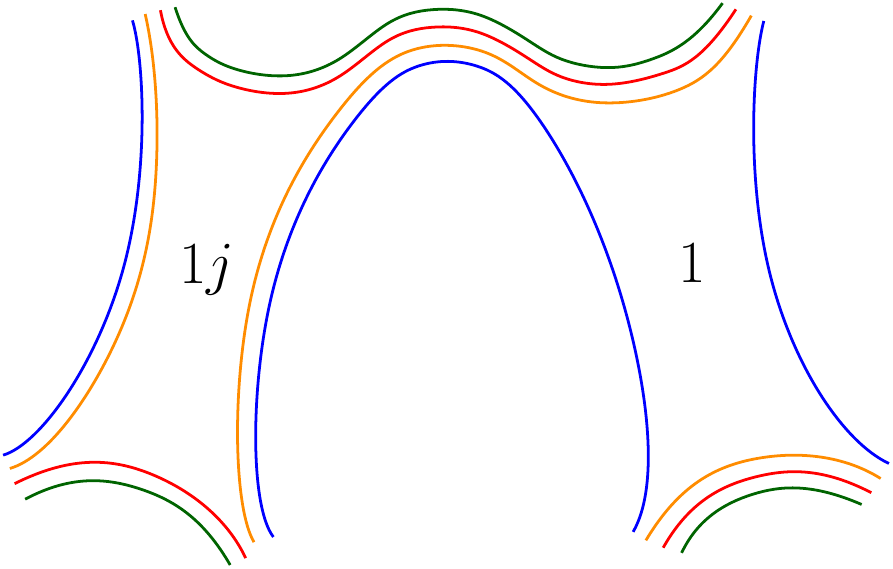}\end{array}
\hspace{0.5cm}\leftrightarrow\hspace{1.3cm}
\begin{array}{c}\includegraphics[scale=1.1]{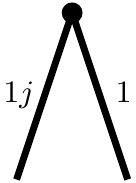}\end{array}
\hspace{2cm}
\ee 
Of course there are several choices of pairings.
If the interaction is of type $\bmp2^i$, 
we always choose to pair the tensors that are linked by 3 strands, and if the interaction is of type  $\bnp2^{1i}$, we always choose to pair tensors that are not linked by a strand of colour 1. 
The colours which propagate between the chosen pairs now label the corresponding edges in the new representation. 
 The vertices in the new representation correspond, in the stranded representation, to cycles alternating pairs of tensors not linked by colour 1 and propagators. In the quartic case, one has the following theorem, first proven in \cite{Bonzom:2015gt}: 
there is a one-to-one correspondence between Feynman diagrams of enhanced four dimensional quartic tensor models and embedded diagrams with edges carrying either one colour $\{i\}$, where $i = 1,2,3, 4$, or a set of two colours $\{1,2\}$, $\{1,3\}$ or $\{1,4\}$. 
Crucially, the faces of colour $i$ of the original Feynman diagrams are mapped to the faces of the embedded colour-$i$ sub-diagram. The total number of faces of an embedded diagram $\cM$ is 
\be
F(\cM)=\sum_{i=1}^4 F_i(\cM).
\ee
Note that if there are no edge with label-set containing colour $i$, then $F_i(\cM)=V(\cM)$. Similarly, if colour 1 appears on every edge, as it is the case if all interactions are of the  $\bnp2^{1i}$ type, then $F_1(\cM)$ is just the total number of faces of the diagram $\cM$. The dependence in $N$ of the amplitude of a diagram is 
\be
\cA(\cM) \sim N^{F(\cM)-\bigl(3\sum_{i=1}^4E_i(\cM) + 2\sum_{i=2}^4E_{1i}(\cM)\bigr)},
\ee
where $E_i$ are edges carrying colour-$i$ only, and $E_{1i}$ are edges labelled with colours 1 and $i$.
Consequently, the embedded diagrams contributing to the leading order of the tensor model are those which maximize $F$ at fixed number of edges of each kind. 
Leading-order contributions are identified by first looking at the behaviour of trees, which are connected embedded diagrams $\cM$ such that $L(\cM)=0$, and then studying the way $F$ changes when deleting edges. One can easily prove recursively on the number of edges \cite{SWMaps} that a tree $\cT$ is such that 
\be
F(\cT)=4+3\sum_{i=1}^4E_i(\cT) + 2\sum_{i=1}^4E_{1i}(\cT),
\ee
corresponding to \eqref{eqref:FTrees}, and resulting in amplitudes scaling as
\be
\cA(\cT) \sim N^4.
\ee
 Furthermore, one can easily show that if a diagram $\cM$ contains an edge labelled with a single colour and which is not a cut-edge (a cut-edge is such that the number of connected components increases by 1 when being deleted), then there exist another diagram $\cM'$ such that 
 \be
\cA(\cM) \sim \cA(\cM') N^{-\alpha},
\ee
 with $\alpha>0$. Thus, $\cM$ cannot contribute to the leading order. 
To identify leading-order diagrams we now use \eqref{eqref:TvsM}. Here, $L_1=L$ and $g_1=g$, so that if $\cM$ and a $\cT$ have the same number of edges of each kind and if colour-$i$ edges in $\cM$ are all cut-edges, the scaling in $N$ of the amplitude is 
\be
\label{eqref:DeltaQuart}
4-\bigl( F(\cT)-F(\cM)\bigr)=4-2L_{m}(\cM)-2\bigl(g(\cM)+\sum_{i=2}^4 g_i(\cM)\bigr),
\ee
where all quantities have been defined before, except 
\be
L_{m}(\cM)=L(\cM)-\sum_{i=2}^4L_i(\cM)\ge 0,
\ee
the number of independent loops in the graph that contain edges of at least two sets of colours $1i$ and $1j$ with $i\neq j$. 
Consequently, a diagram contributing to the leading order is planar ($g(\cM)=g_i(\cM)=0$), and only has loops of edges that carry the same two colours 1 and $i$ ($L_m=0$) for any $i$.  A schematic example is shown in (\ref{fig:Cactus}) (each blob corresponds to a planar component with only edges carrying the same colour-set $\{1i\}$).
\be
\label{fig:Cactus}
\begin{array}{c}\includegraphics[scale=0.5]{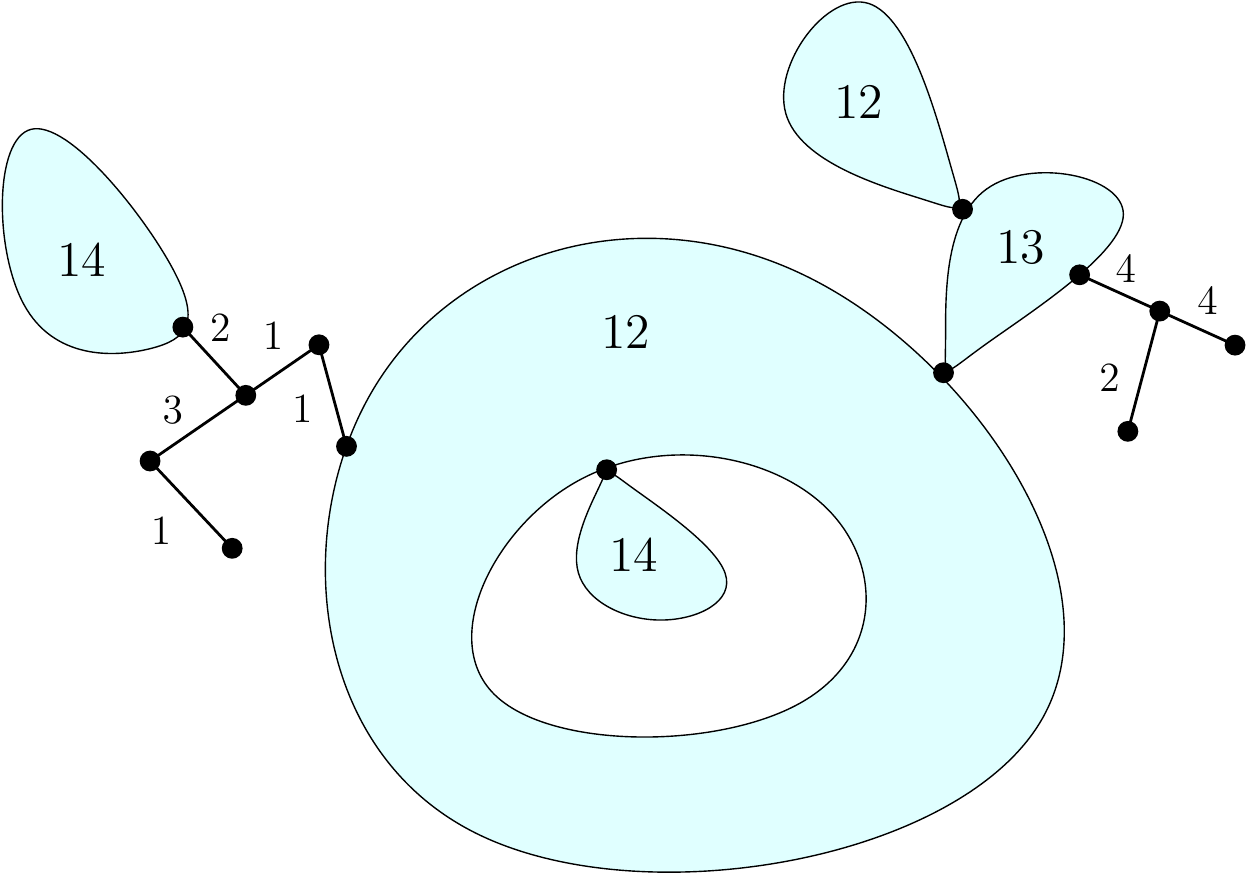}\end{array}
\ee
%

We typically count diagrams with one distinguished edge, to which an orientation is given. The reason is that this  breaks symmetries, as mentioned previously, and therefore leading order diagrams can then be counted more easily. Cutting that distinguished edge, one obtains a contribution to the 2-point function, and conversely. Counting rooted diagrams therefore goes back to counting leading-order contributions to the 2-point function.
%

Non-separable embedded diagrams \cite{Goulden:2004uj} are  such that no face has two corners incident to the same vertex.
The distinguished edge is either a cut-edge of colour $i$, in which case there is one 2-point function at each extremity, and the contribution is $t_i\gt^2$, or it belongs to a leading-order non-separable planar component, in which case all the edges have the same colour set $\{1i\}$. In each corner on each vertex is a (possibly trivial) rooted leading-order 2-point function.
We denote $\nsgf$ the generating function of rooted non-separable planar embedded diagrams, 
\be
\nsgf(z)=\sum_{n\ge 0} c_n z^n,
\ee
where $c_n$ is the number of non-separable embedded diagrams with $n$ edges. 
It is given by the following parametrization \cite{Tutte:63, Goulden:2004uj}, 
\bea
\label{eqref:nonsepGF}
z&=&u(1-u)^2 \\
\nsgf(z)&=&(3u+1)(1-u).
\eea
In an embedded diagram, the number of corners is twice the number of edges. 
Because we need to add one 2-point function $\gt$ per corner of the non-separable component containing the root, we need to replace the parameter $z$ counting edges of the non-separable planar component with $\tnp{2}^{1i}\gt^2$. We therefore obtain the following equation for the leading-order two-point function $\gt$, which depends on the coupling constants $\{\tmp2^i\},\{ \tnp2^{1i}\}$
\be\label{eqref:DSEnecklace}
\gt = 1 + \sum_{i=2}^4 \bigl( \nsgf(\tnp2^{1i} \gt^2)-1\bigr) + \gt^2\sum_{i=1}^4 \tmp2^i.
\ee

The reduction of this equation to the case with a single necklace and melonic interactions  has been investigated in \cite{Bonzom:2016to}, as well as the case where all necklaces are given the same coupling. In Section~\ref{sec:MultiCrit}, we study in detail the critical regimes when considering quartic necklaces with independent couplings, and show that the same critical regimes are involved.

\subsection{The sextic case in dimension three}
\label{subsec:K33LO}

For $\ktt$-type interactions the total number of faces of leading-order Feynman diagrams is
\be
\label{eqref:TreeK33}
F(\cM)=3(1+\B(\cM)),
\ee
as shown in \cite{SWMaps} where we recall that $\B$ is the number of interaction bubbles, so that when putting a $N^3$ scaling in the action, corresponding to $s=1$, the amplitude of leading-order Feynman diagrams contributing to the free energy scales as $N^3$.
Vacuum leading-order diagrams are such that bubbles are locally as in one of the two situations illustrated in \eqref{fig:K33LO}, as proven in \cite{SWMaps},
\be
\label{fig:K33LO}
\begin{array}{c}\includegraphics[scale=0.5]{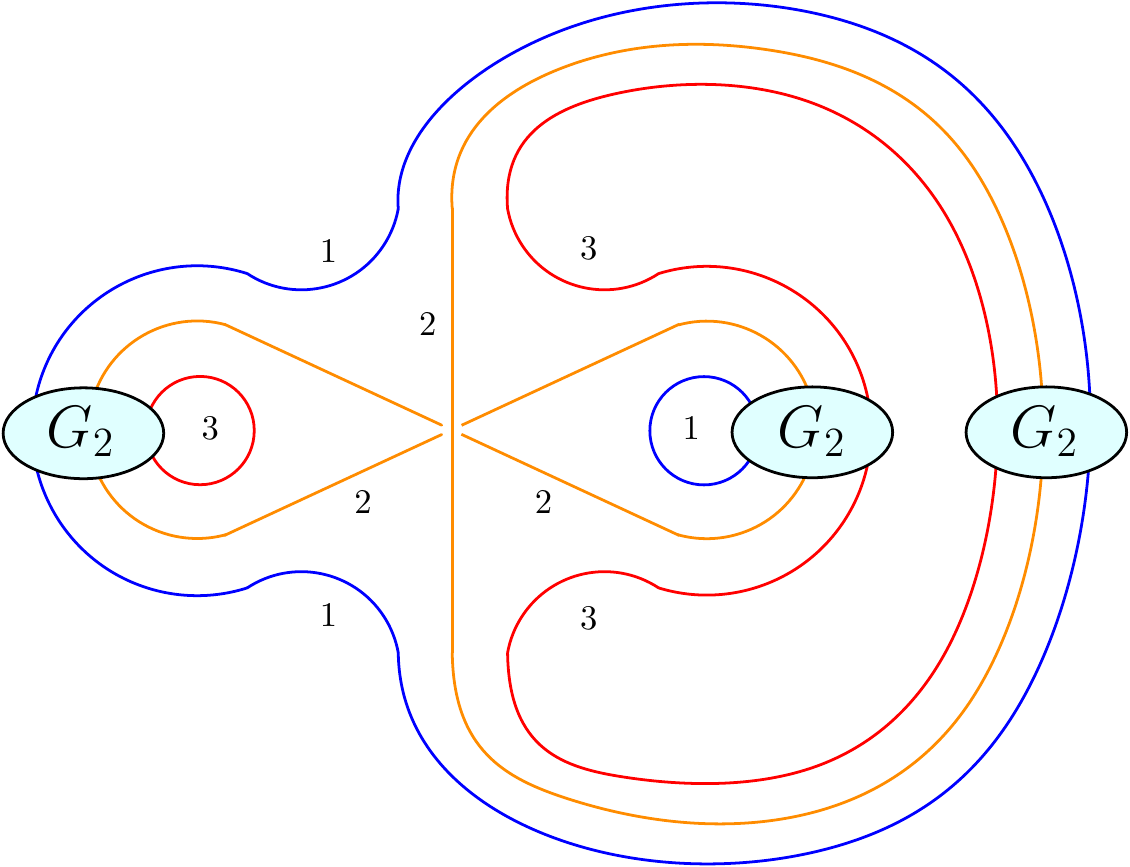}\end{array}
\hspace{2cm}
\begin{array}{c}\includegraphics[scale=0.4]{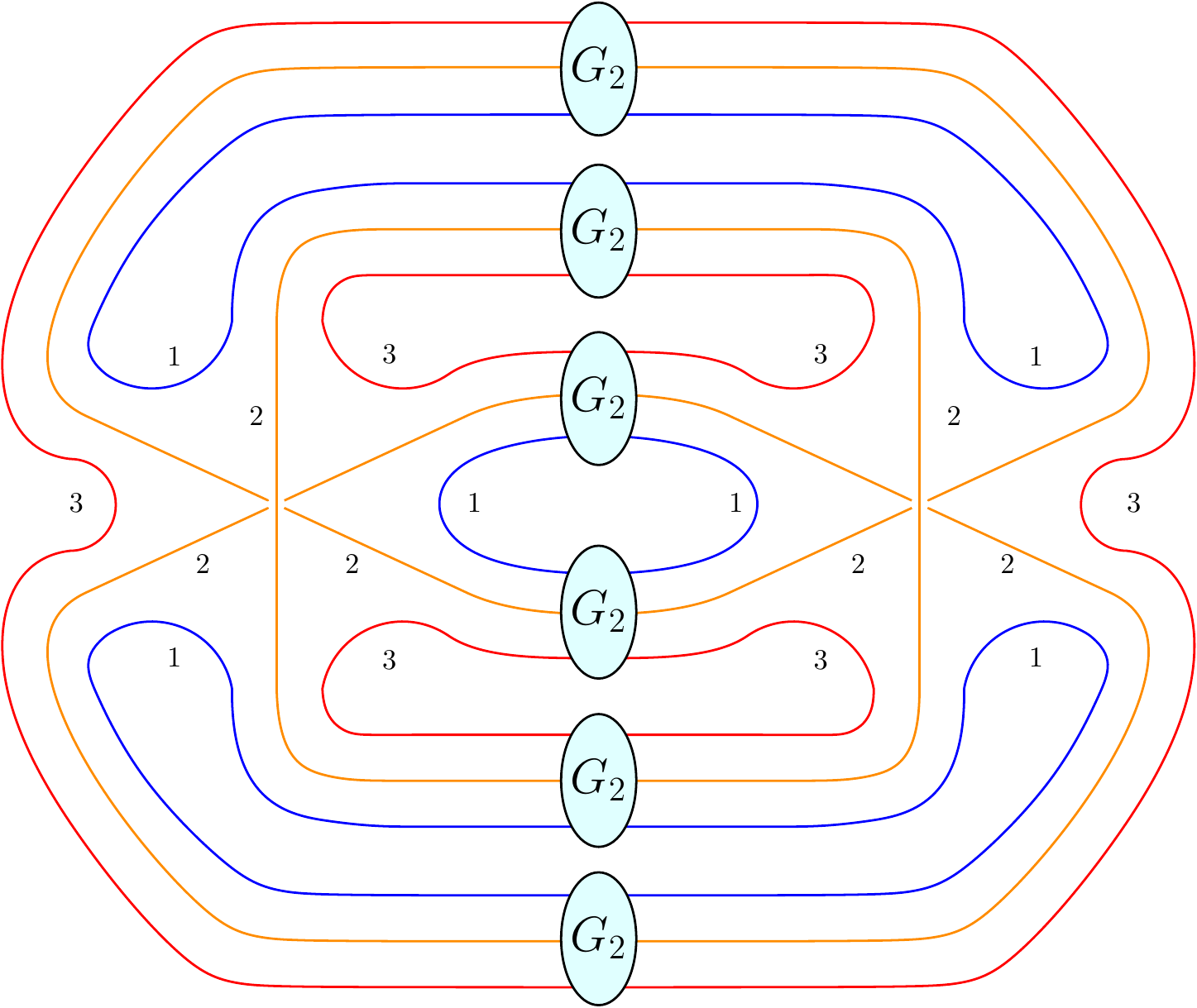}\end{array}
\ee
in which the $\gt$ blobs indicate the insertion of a possibly trivial contribution to the leading-order 2-point function. 
As explained previously, we count the diagrams contributing to $\gt$. It corresponds to counting diagrams with one distinguished edge, so that all symmetries break down. Distinguishing an edge goes back to choosing a particular pair $(T, \bar T)$.
On each side, given a $T$, there are three different ways of choosing a $\bar T$.
Consequently, the self-consistency equation of the 2-point function $G_2$ is
\be 
\label{eqref:GenFunK33}
\gt(\tk)=1+3\tk \gt\bigl(\tk)^3+ 3\tk^2\gt(\tk)^6.
\ee 
We stress that the self-consistency equation for the 2-point function of all sextic melonic interactions would be the same, but without the order 6 term. The effect of the non-planarity is therefore to generate loops at leading order, but which are too small to change the large-scale structure of leading-order diagrams.

For this model, we introduce a new correspondence between Feynman stranded diagrams and embedded diagrams, subtler  than that described at the beginning of Sec.~\ref{sec:LOdiagrams} and in the Appendix,  which renders the tree-structure of leading-order diagrams explicit at the cost of introducing extra colours on the (white) intermediate-field vertices corresponding to the interaction bubbles. 
This additional colouring takes into account the way the tensors $T$ and $\bar T$ have been paired to obtain the patterns of (\ref{fig:K33LO}).\footnote{More precisely, the correspondence is similar to that described in the Appendix, but instead of choosing one given pairing of the $(T, \bar T)$ for all the interactions in the graph, we choose the pairing locally for every subgraph as in \eqref{fig:K33LO}, making sure that the paired tensors are those linked by a propagator with a $G_2$ blob. 
For technical details see \cite{Lionni:17}, Sec.~3.2.6.} Leading-order stranded diagrams are mapped to embedded diagrams with coloured vertices and black vertices such that edges - which carry colour sets $\{12\}$, $\{13\}$, $\{23\}$, or $\{123\}$ - can only link coloured vertices to black vertices. 
The black vertices are of any valency and are similar to the vertices introduced in Section~\ref{subsec:Quart}. The other vertices are of valency 3 (colours $a$, $b$ and $c$), or 6 (colours $d$, $e$, and $f$), and are \textit{rigid} in the sense that they have a fixed ordering of edges around them. The cyclic counterclockwise ordering of edges around valency-3 vertices $a$, $b$ and $c$, is given by their colour-sets according to $\{12\}$, $\{13\}$, $\{23\}$,  as shown in (\ref{fig:K33LOBij1}). 
\be
\label{fig:K33LOBij1}
\begin{array}{c}\includegraphics[scale=0.5]{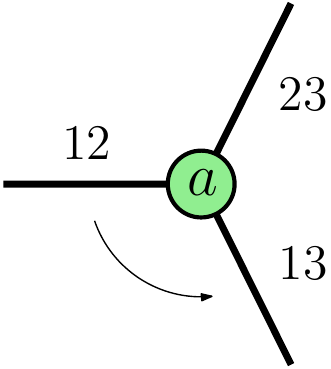}\end{array}
\hspace{2.6cm}
\begin{array}{c}\includegraphics[scale=0.5]{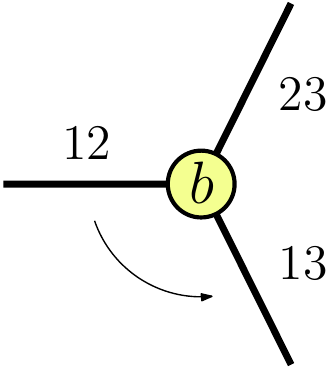}\end{array}
\hspace{2.6cm}
\begin{array}{c}\includegraphics[scale=0.5]{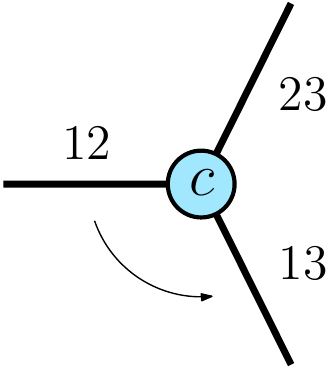}\end{array}
\ee
The 6-valent vertices $d$, $e$, and $f$ have a coloured internal structure, and all incident edges carry all colours $\{123\}$, as shown in (\ref{fig:K33LOBij2}).
\be
\label{fig:K33LOBij2}
\begin{array}{c}\includegraphics[scale=0.6]{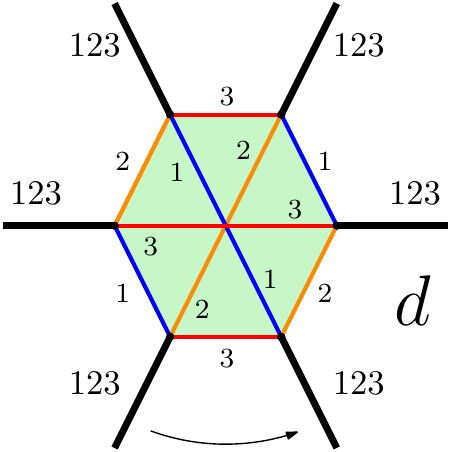}\end{array}
\hspace{1.5cm}
\begin{array}{c}\includegraphics[scale=0.6]{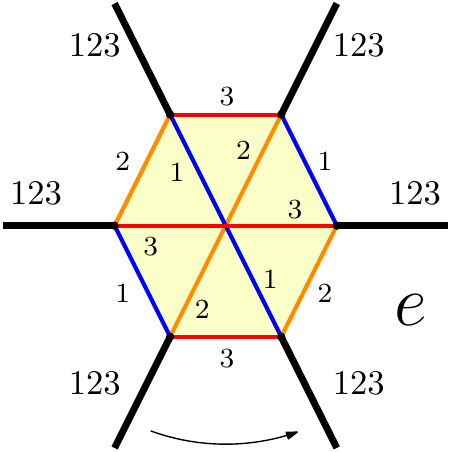}\end{array}
\hspace{1.5cm}
\begin{array}{c}\includegraphics[scale=0.6]{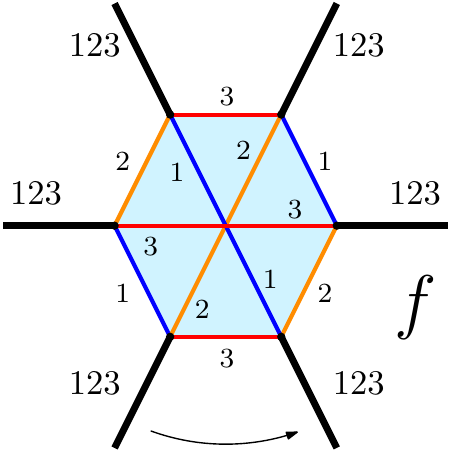}\end{array}
\ee
 As before, the faces of colour $i$ of the stranded diagram in the original picture are now mapped to the faces of the corresponding colour-$i$ sub-diagram. Leading-order diagrams are mapped to trees in the new representation, and other diagrams are mapped to embedded diagrams as just described, which have at least one loop and have a few forbidden small sub-diagrams.\footnote{In order to stay concise, we do not explain in detail the correspondence for non-leading diagrams, as we only use the correspondence for leading-order diagrams in the following. We stress however that this correspondence can be extended to all orders.}
The number of faces of a tree in the new representation is recursively proven to be (\ref{eqref:TreeK33}), which corresponds to equation \eqref{eqref:FTrees} and provides $a_{3_\textrm{K}}=3$, and therefore $s_{3_\textrm{K}}=1$ from \eqref{eqref:AfromS}. As in the previous subsection, we define $L_{\overline{12}}\ge0$, the number of independent loops of $\cM$ which are not counted in $L_1$ and $L_2$, as
\be
\label{eqref:L12A}
L_{\overline{12}}= L- L_1 - L_2.
\ee
In particular, we will use the two following properties:
\be
\label{eqref:L12B}
L\ge L_i+L_j
\ee
for any $i\neq j$, and 
\be
\label{eqref:L12C}
 L_{\overline{ij}}(\cM)>0\quad \Rightarrow \quad \biggl[\,\substack{{\textrm{There exist some edge in } \cM \textrm{ which is a cut-edge in }}\\[+1ex]{ \cM_i {\rm\ and\ } \cM_j \textrm{ but is not a cut-edge in } \cM}}\,\biggr].
\ee
Using Relation~(\ref{eqref:TvsM}) which compares the number of faces of intermediate-field diagrams and trees, we see that a diagram $\cM$ satisfies
\be
F(\cM)=3(1+\B(M))-\Delta_3(\cM)=0,
\ee
where $\Delta_3(\cM)$ is the quantity 
\be
\label{eqref:Delta3}
\Delta_3(\cM)=L_{\overline{12}}+L_{\overline{13}}+L_{\overline{23}} + 2\sum_{i=1}^3g_i(\cM)\ge0,
\ee
which is non-negative as a sum of positive quantities, and vanishes iff all terms vanish. The amplitude of diagrams in the new representation is 
\be
\cA(\cM)\sim t_K^{\B(\cM)} N^{3-\Delta_3(\cM)},
\ee
where the number of bubbles $\B(\cM)=V_{\rm col}(\cM)$ is the number of coloured vertices and we can now restate the condition a diagram needs to satisfy in order to contribute at leading order,
\be
\cM \text{ is L.O.} \quad \Leftrightarrow \quad \bigg\{\begin{array}{rcccccl} L_{\overline{12}}&=&L_{\overline{13}}&=&L_{\overline{23}}&=&0 \\
g_1&=&g_2&=&g_3&=&0
\end{array}.
\ee

This is the characterization used in \cite{SWMaps, Lionni:17} to prove that the leading order diagrams are as shown in \eqref{fig:K33LO}. The proof is quite lengthy, so we do not repeat it here, but just sketch it: The leading order embedded diagrams are characterized recursively on the number of white vertices. The leading order diagrams with one and two vertices are identified to be the embedded diagrams corresponding to the diagrams of \eqref{fig:K33LO} but without the $G_2$ blobs. Then it is recursively shown that if $\Delta_3(\cM)=0$, then $\cM$ contains a subgraph as in \eqref{fig:K33LO}, and the blobs still satisfy $\Delta_3(\cM)=0$, so the recursion hypothesis applies to them. For more details, we refer the reader to \cite{Lionni:17}.

\subsection{The sextic case in dimension four}
\label{subsection:K334}

We explain now the leading-order behaviour of all $\rk=4$ sextic interactions.
As mentioned in Subsection~\ref{subsec:Ord6Rg4Intro}, the interactions of type $3_\textrm{M}$ and $3'_\textrm{M}$ in Table~\ref{table:rank4-interactions} are melonic, and therefore their critical behaviour and continuum limit is known. The tensor model built with necklace interactions of type $3_\textrm{N}$, as well as melonic interactions $3_\textrm{M}$ and $3'_\textrm{M}$  is similar to the full quartic rank-4 tensor model in all aspects, and the proof follows the exact same steps. Trees belong to the leading order and behave as \eqref{eqref:FTrees} with number of faces
\be
F^{3_\textrm{N}}_\text{LO}= 4 + 4\B_{3_\textrm{N}} + 6(\B_{3_\textrm{M}} + \B_{3'_\textrm{M}})  
\ee
such that $a_{3_\textrm{N}}=4$, and therefore $s_{3_\textrm{N}}=2$ from \eqref{eqref:AfromS}. Denoting $\nsgf^{(3)}(z)$ the generating function of planar ribbon graphs with white vertices of valency three and black vertices of any valency, and such that no face visits twice or more the same black vertex, counted according to their number of black vertices\footnote{This can be computed from \cite{Bender:1994kg}.}, the self-consistency equation for the leading-order 2-point function of this model is
\be
\gt = 1 + \sum_{i=2}^4 \bigl( \nsgf^{(3)}(\tnp3^{1i} \gt^3 )-1\bigr) + \gt^3 \bigl(\sum_{i=1}^4 \tmp3^i+ \sum_{i<j=1}^4 t_{3'_\textrm{M}}^{ij}\bigr)
\ee
as $\gt$ is inserted on every corner around a black vertex, and there are as many such corners as edges in the diagram. We expect the same critical regimes as in the quartic case.

\emph{Interactions of type} $3_\textrm{MN}$ are obtained from a quartic necklace by ``opening" a strand of colour $i$ and inserting two tensors linked by three new strands of colours different from $i$. These two tensors are paired, and, depending on whether colour $i$ is colour 1 or parallel to colour 1, or not, we have the following intermediate-field vertices (with the pairing rule (\ref{fig:Pairings})),
\be 
\begin{array}{c}\includegraphics[scale=0.3]{diagrams/NeckMelS.pdf}\end{array} \quad\leftrightarrow\quad
\begin{array}{c}\includegraphics[scale=0.5]{diagrams/NeckMelIF.pdf}\end{array},
\qquad
\begin{array}{c}\includegraphics[scale=0.3]{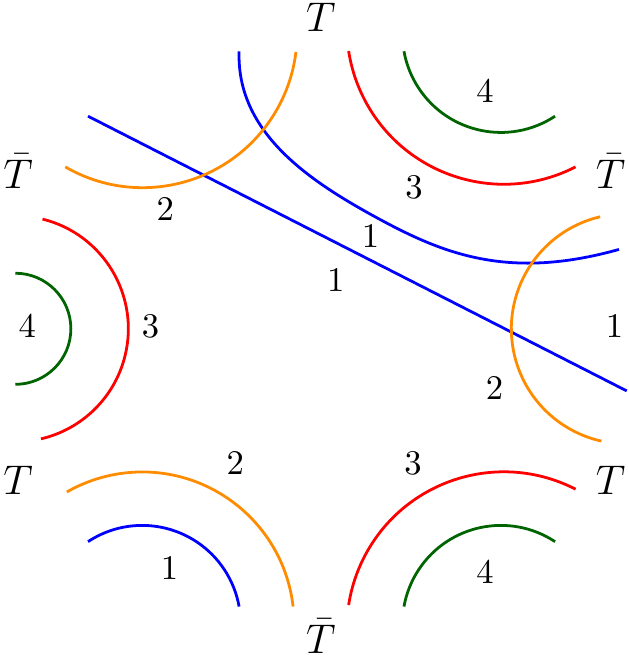}\end{array} \quad\leftrightarrow\quad
\begin{array}{c}\includegraphics[scale=0.5]{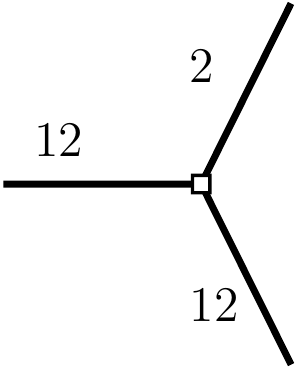}\end{array},
\qquad
\begin{array}{c}\includegraphics[scale=0.3]{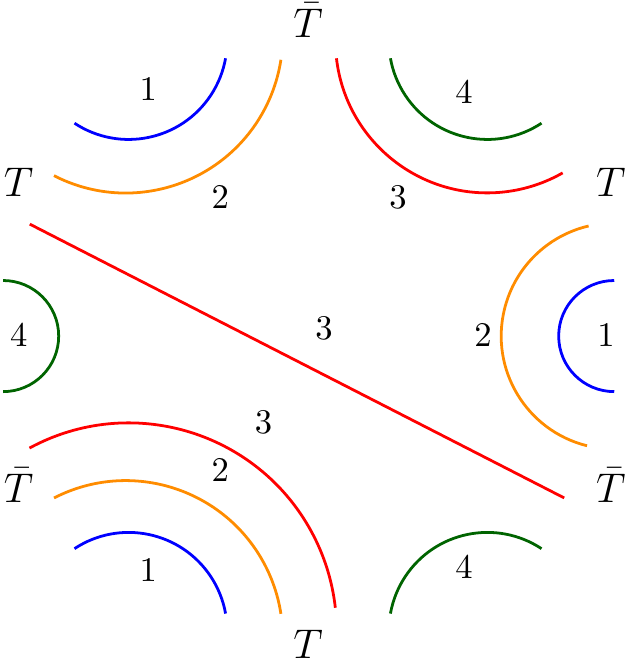}\end{array} \quad\leftrightarrow\quad
\begin{array}{c}\includegraphics[scale=0.5]{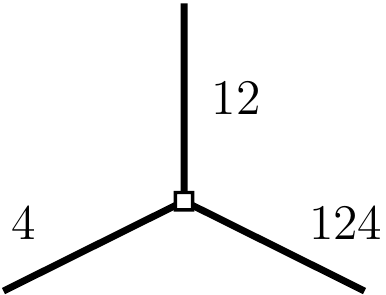}\end{array},
\ee 
and their equivalents. The edge corresponding to the insertion always carries a single colour. As in the quartic case, such an edge is necessarily a cut-edge in every leading-order diagram. Indeed, unhooking an edge $e$ as shown in (\ref{fig:Unhook}) (here, we cannot delete an edge because of the rigid structure of white vertices),
\be
\label{fig:Unhook}
\begin{array}{c}\includegraphics[scale=0.6]{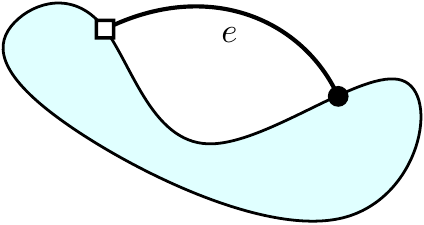}\end{array}
\qquad\rightarrow\qquad
\begin{array}{c}\includegraphics[scale=0.6]{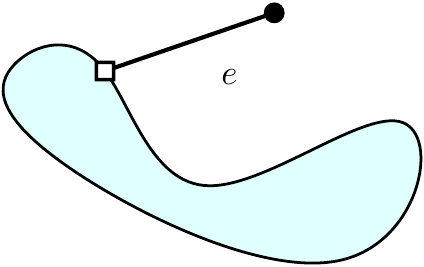}\end{array}
\ee
 it is easily seen that, if $e$ carries a single colour, the number of faces necessarily increases at least by two. Therefore, if unhooking the edge does not increase the number of connected components, the diagram is suppressed at large $N$. Leading-order diagrams of a type $3_\textrm{MN}$ tensor model are obtained from leading-order diagrams from necklace-type models by performing the following operations on the edges
\be 
\begin{array}{c}\includegraphics[scale=0.5]{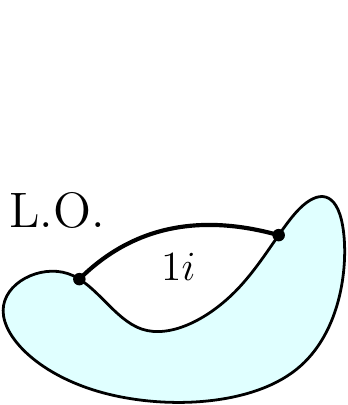}\end{array} \qquad\rightarrow\qquad
\begin{array}{c}\includegraphics[scale=0.6]{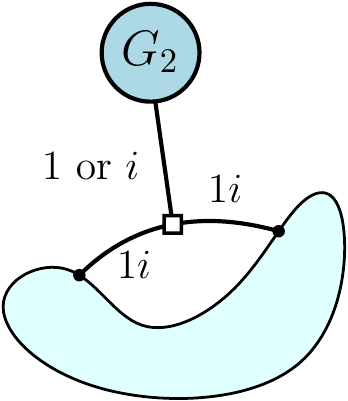}\end{array}
\qquad\text{or}\qquad
\begin{array}{c}\includegraphics[scale=0.6]{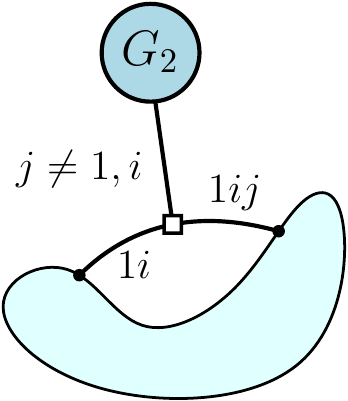}\end{array},
\ee 
where $\gt$ is a 2-point leading-order insertion from the current order-6 model. Consequently, trees belong to the leading order, and it is seen recursively that their number of faces behave as
\eqref{eqref:FTrees} with
\be
F^{3_\textrm{MN}}_\text{LO}= 4 +5\B, 
\ee
yielding $a_{3_\textrm{MN}}=5$, and therefore $s_{3_\textrm{MN}}=1$ from \eqref{eqref:AfromS}. 
A distinguished edge in a leading-order diagram is always incident to a non-separable planar diagram with all edges of colour $1i$, with one full $\gt$ insertion on every corner, and another $\gt$ inserted on each edge, with 3 ways of deciding which one is the distinguished edge. For each edge, one has to decide which one of the four colours $c$ of the quartic necklace $\bnp2^{1i}$  has received an insertion, contributing with a coupling $t_{3_\textrm{MN}}^{ci}$. The self-consistency equation for the 2-point function of the full type-$3_\textrm{MN}$ tensor model is 
\be
\gt = 1 + \sum_{i=2}^4\bigl\{ \nsgf\bigl( 3\gt^3\sum_{c=1}^4 t_{3_\textrm{MN}}^{ci}\bigr)-1\bigr\}.
\ee
In the case where a single interaction $\cB^{ci}_{3_\textrm{MN}}$, with $c\in\{1,2,3,4\}$ and  $i\in\{2,3,4\}$, is allowed it reduces to the simpler equation
\be
\gt =  \nsgf\bigl( 3\gt^3 t_{3_\textrm{MN}}^{ci}\bigr).
\ee

We now consider an \emph{interaction of type} $3'_\textrm{K}$, and show that the leading-order intermediate-field diagrams are precisely trees. 
Suppose that 4 is the colour which belongs to every double-strands, as in the diagram of Table~\ref{table:rank4-interactions}. We apply relation \eqref{eqref:TvsM}, which compares faces of trees $\cT$ and diagrams $\cM$ with the same number of interactions: 
\bea
F(\cT) -F(\cM)&=& L(\cM) -2 L_4(\cM)+ 2g_4(\cM)+ \Delta_3(\cM)\\
&\ge&  L(\cM) -2 L_4(\cM)+ 2g_4(\cM),
\eea
where $\Delta_3$ has been defined in \eqref{eqref:Delta3}. As colour 4 does not appear on any edge, $L_4(\cM)=g_4(\cM)=0$ such that 
\be
F(\cT) - F(\cM) = L(\cM)   .
\ee
This vanishes if $\cM$ is a tree and is positive otherwise. The self-consistency equation is a simple cubic tree-equation, and, as for $3_\textrm{MN}$ type interactions, $a_{3'_\textrm{K}}=5$, and $s_{3'_\textrm{K}}=1$. This argument holds for a single type $3'_\textrm{K}$ colouring, however we also expect a tree-behaviour for the full $3'_\textrm{K}$ tensor model.

\

We consider a single \emph{interaction of type} $3_\textrm{K}$, with the colouring as in Table~\ref{table:rank4-interactions}. The number of faces of trees is \eqref{eqref:FTrees}
\be
F(\cM)=4(1+\B(\cM)).
\ee
Moreover, diagrams are such that $L_3=L_4$ and (\ref{eqref:L12A}, \ref{eqref:L12B}, \ref{eqref:L12C}) are true for $i\neq j <4$,  so that  (\ref{eqref:TvsM}) rewrites as 
\be
\label{eqref:Delta4}
\Delta_4(\cM)=F(\cT)-F(\cM)= 2(L_{\overline{13}}(\cM)+L_{\overline{23}}(\cM)) + 2\sum_{i=1}^4 g_i(\cM),
\ee
so that trees belong to the leading order, from which we deduce that, as $a_{3_\textrm{K}}=4$,  $s_{3_\textrm{K}}=2$ from \eqref{eqref:AfromS}, and other leading-order contributions are such that 
\be
\label{eqref:CondK334}
L_{\overline{13}}=L_{\overline{23}}=0, \textrm{ and } \forall i,\ g_i=0.
\ee
There are two cases. If in addition $L_{\overline{12}}=0$, then we know from the previous section that in our representation, the diagram is a tree. If not, $L_{\overline{12}}(\cM)>0$, and from (\ref{eqref:L12C}) we know that there exist an edge $e$ which is not a cut-edge, but is so in both $\cM_1$ and $\cM_2$. Unhooking $e$ as in (\ref{fig:Unhook}),
we obtain a new diagram $\cM'$ such that,  
\be
L'=L-1,\qquad
L_1'=L_1,\quad \text{and}\quad
L_2'=L_2.
\ee
Using (\ref{eqref:L12B}) and (\ref{eqref:CondK334}), 
\be
\label{eqref:L13}
0\le L'-L_1'-L_3'=L-1-L_1-L_3'=L_3-L_3'-1.
\ee
Furthermore,  $L_3-L_3'$ is 0 or 1, but the first value contradicts (\ref{eqref:L13}), so that 
\be
 L_3'=L_3-1,
 \ee
and therefore $\cM$ is obtained from $\cM'$ by hooking an edge of colour $i34$ to create a loop such that $\cM_3=\cM_4$ stays planar and such that it is a cut-edge in $\cM_i$. We still have $L'\ge L_1'+L_2'=L_1+L_2$, so there are two cases. Either $L'> L_1'+L_2'$, in which case we repeat the same steps, or $L'= L_1'+L_2'$, in which case $\cM'$ is a tree. From this tree, $\cM$ is obtained by hooking a certain number of edges $i34$ so that every edge of colour $12$ stays a cut-edge and such that the overall diagram remains planar.

Leading-order diagrams are quite similar to those obtained from a single type $3_\textrm{MN}$ interaction, with the difference that here, the tree part has three valency-3 vertices and three valency-6 vertices.  We distinguish an edge with colour set $\{134\}$ or \{234\}. 
To recover the leading-order 2-point function, one simply replaces $z$ with $3/2 z$.
%
 There are two cases. 
 In the first case, the distinguished edge is either a cut-edge of colour set $\{134\}$ or $\{234\}$, and it belongs either to one of the three 3-valent vertices or to one of the three 6-valent vertices. The contribution is $6 \tk\gt(\tk)^3  + 6 \tk^2 \gt(\tk)^6$. 
 In the other case,  the distinguished edge is not a cut-edge, in which case it belongs to a non-separable planar component \eqref{eqref:nonsepGF} for which all edges contain the colours 34, and we need to insert one 2-point function per black corner of the non-separable component and one 2-point function per edge - corresponding to the colour-$\{12\}$ cut-edge -  that can be added in two possible ways. Indeed, a white vertex is added on every edge, and a colour-$\{12\}$ bridge is added between that white square and a leading order sub-diagram, and there are two ways of deciding which one of the two colour-$\{34\}$ segments has colour-set $\{234\}$, and which one has colour-set $\{134\}$. This second possibility contributes as $\nsgf\left(2 z \gt ^3\right)$. It generates a tree part $2z\gt^3$ which has to be subtracted from the overall tree part.
The 2-point function $\gt$ therefore satisfies 
\be\label{eqref:DSE-K334}
\gt = \gt(\tk) = \nsgf \left(2 \tk \gt ^3\right)+4 \tk \gt^3  + 6 \tk^2 \gt^6. 
\ee
 
As for interactions of type $3'_\textrm{N}$, we notice that a propagator can be inserted on every interaction bubble,
\be 
\begin{array}{c}\includegraphics[scale=0.4]{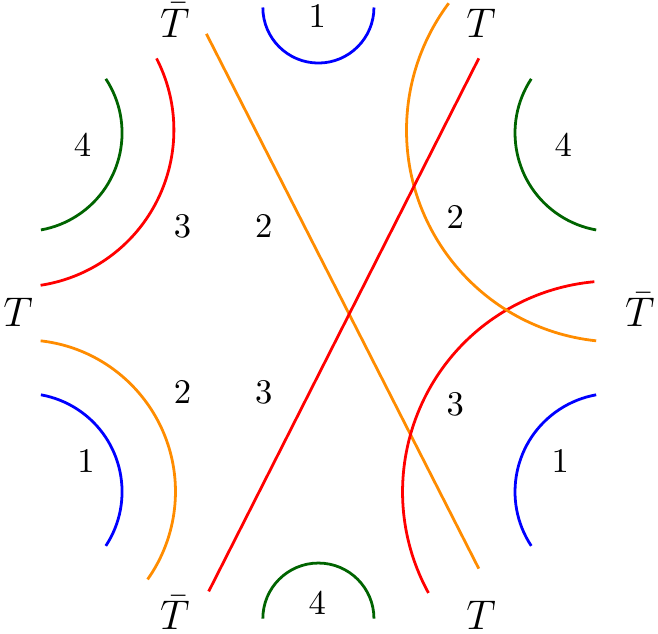}\end{array} \qquad\rightarrow\qquad
\begin{array}{c}\includegraphics[scale=0.4]{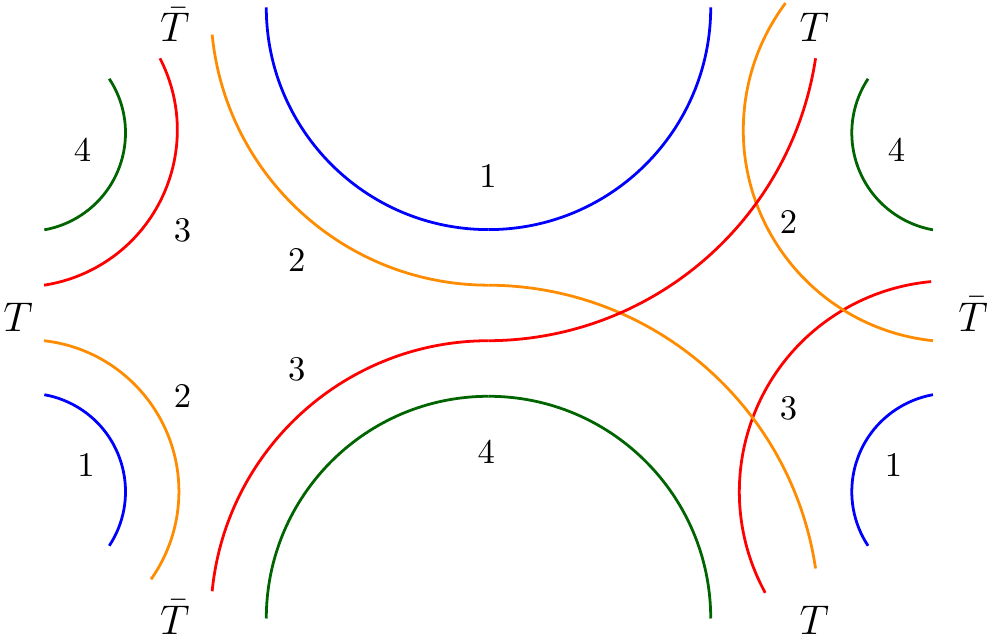}\end{array},
\ee 
transforming every bubble into two quartic necklaces with different colouring and a distinguished propagator between them, without changing the number of faces. 
A type $3'_\textrm{N}$ diagram belongs to the leading order if the diagram obtained by inserting a propagator for every bubble is a type-$2_\textrm{N}$ leading-order diagram. Because the inserted propagators are distinguished, leading-order diagrams of type $3'_\textrm{N}$ correspond to leading-order diagrams of type $2_\textrm{N}$ for which there exist a way of choosing propagators such that every bubble is incident to a distinguished propagator, and such that every distinguished propagator is between two quartic necklaces with different colourings. 

This is a very specific kind of hard-dimer model on leading-order quartic necklaces (it is similar to a dimer model on random lattices with black and white vertices such that distinguished edges only link black and white vertices, and such that every  vertex is incident to a distinguished edge). 
Coupling a model to hard dimers is in general similar to considering multi-critical behaviour for models with higher order interactions. This suggests that the critical behaviour from interactions of type $3'_\textrm{N}$ could in principle be obtained from the quartic necklaces regimes. 
Each one of the various models involving interactions of type $3'_\textrm{N}$ and  $2_\textrm{N}$ should correspond to a very specific kind of hard-dimer model on random lattices with coloured vertices and particular restrictions on how the distinguished edges can be, which could exhibit very interesting critical behaviour. 
We will however leave their study for further work.


\section{Multi-critical behaviour of the models}
\label{sec:MultiCrit}

Enhanced coloured tensor models exhibit a rich multi-critical structure at leading order in the $1/N$ expansion.
Critical behaviour of branched-polymers is the standard case in tensor models, but enhancing necklace-type interactions allows for planar regimes as well. When both types of interactions are present in the action, there are regimes in coupling space for each behaviour.
The boundary between them obeys the special transition behaviour for which ``cacti" dominate which are made of an infinite number of discrete spheres of diverging areas - baby universes- with tree-like incidence relations. These are characterized by a critical exponent $\gamma=1/3$.
Moreover, for several couplings the regimes have further internal structure of multi-critical subspaces. 
In this section we explain this structure for the models under consideration and illustrate it with explicit examples. 
As a new result we find that there is generically no multi-criticality on the boundary between the regimes induced by trees.

\subsection{Universality transition between planar and branched-polymer regimes}

All the models under consideration yield equations for the two-point function $\gt$ of the general form
\be\label{eqref:DSEgeneral}
\gt = Q_0[\gt] + \sum_j \nsgf(Q_j[\gt])
\ee
where $Q_0, Q_1,...$ are a finite number of polynomials in $\gt$ and $\nsgf$ is the generating function of non-separable planar diagrams, \eqref{eqref:nonsepGF}.
(Strictly speaking, sextic necklaces $\bnp3$ and the necklace-type interaction $\cB_{3'_\textrm{N}}$ have to be described by a modified generating function $\nsgf^{(3)}$ for non-separable rooted diagrams with restricted valency, or dually, faces restricted to hexagons. Nevertheless, there are no reasons to expect qualitative differences.)

\

We conjecture that solutions $\gt$ always have the same qualitative behaviour of a planar regime, a branched-polymer regime and a baby-universe transition phase.
To this end we have analysed the critical behaviour of the various models considered here.
We provide some general arguments and illustrate this statement with a number of explicit examples in the following parts of the section.

Let us consider the general case of one planar coupling $\tp $ and a finite number of melonic couplings $\{\tm j\}$. 
The corresponding equations for $\gt$  are
\be\label{eqref:DSEmelons+necklace}
0 = \lop(\gt,\tm j) + \nsgf( \tp \gt^k)
\ee
where $V$ covers a finite set of melonic interactions,
\[\label{eqref:lopotential}
\lop = \lop(\gt,\tm j) = -\gt + \sum_{j\ge 1} \tm j \gt^j\, ,
\]
and $k\ge2$ results from the interaction of order $2k$ that is relevant for the planarities. Here, $k=2$ for quartic necklaces \eqref{eqref:DSEnecklace}, and $k=3$ for type $3_\textrm{K}$ interaction \eqref{eqref:DSE-K334}.

Heuristically, the generic two-phase structure of the models is obvious from \eqref{eqref:DSEmelons+necklace}: For the planar coupling $g$ much larger than the melonic couplings $\{\tm j\}$, the $V$ part can be neglected and the equation is simply the equation for planar diagrams \cite{Bonzom:2016to}. On the contrary, for $g$ much smaller than the $\{\tm j\}$, the equation reduces effectively to $0= \lop(\gt,\tm j) $ for which the implicit-function scheme theorem \cite[Ch.\,VII.4]{Flajolet:2009wm} applies yielding critical branched-polymer behaviour.

\

We have used three methods to determine critical loci.
Due to the definition \eqref{eqref:nonsepGF} of $\nsgf$, the equations for $\gt$ are parametric. The straightforward way to determine critical loci is given by the vanishing of the Jacobian determinant together with the original equations. This is again a system of equations which has to be solved for $\gt$, the additional parameter $u$ (see \eqref{eqref:nonsepGF}) and one of the couplings. The advantage is that all critical values including $\gt^\cri$ are directly determined.

A second way to find the critical loci is to reduce the parametric equations to a single one and demand its discriminant to be vanishing.
For \eqref{eqref:DSEmelons+necklace}, elimination of the extra variable of the parametric definition of $\nsgf$ \eqref{eqref:nonsepGF} in terms of the resultant yields the equation
\[
0 = \lop^2 (\lop+1) -2 (8 + 9 \lop) \tp \gt^k -27 (\tp \gt^{k})^2 \, .
\]
Up to some monomial factors, its discriminant has the form of a product $ P_\bp\cdot(P_\pla)^3$ of a polynomial $P_\bp$ determining the surfaces at which $\gt$ obeys the critical behaviour of branched polymers and $P_\pla$ yielding planar critical behaviour. The transition line between the two regimes is then determined by the common solution $P_\bp = P_\pla = 0$.

A third method is to determine critical loci of $\gt(\tp,\tm j)$ as extrema of the inverse function $\tm k(\gt,\tp,\tm j)$ for some $\tm k\in\{\tm j\}$. While this is straightforward when equations are linear in the couplings, for polynomial equation like in our case a clever rescaling of couplings has to be guessed in order to find an explicit inversion, if possible at all.
Indeed, \eqref{eqref:DSEmelons+necklace} can be appropriately inverted after rescaling $\tm j = \lm j t$ and $\tp = \lp t$. Then $\lop(\gt,\tm j) = t \lopt(\gt,\lm j)$ and the equation takes the form
\[
0 = \lopt^3 t^2 + (9 \lp \gt^k - \lopt)^2 t - 108 (\lp\gt^{k})^2 t - 16\lp \gt^k \,.
\]
such that
\[
t_\pm = t_\pm[\gt, \lp, \lopt(\gt,\lm j)] = \frac1{2 \lopt^3} \left[(9 \lp \gt^k - \lopt)^2 - 108 (\lp\gt^{k})^2 \pm (\lp \gt^k - \lopt)^{\frac1 2}(9 \lp \gt^k - \lopt)^{\frac3 2} \right] \, 
\]
where $t_{-}$ is the physical branch with $t_{-}(\gt=1,\lp,\lopt)=0$ .

The third method is particularly useful to determine multi-critical loci with exponents
\[
\delta \equiv 1 -\gamma = \frac 1{q+1} \quad \rightarrow \quad \gamma  = \frac q{q+1}
\]
from vanishing derivatives
\[
0 = \frac{\partial t}{\partial \lopt} = \frac{\partial^2 t}{\partial \lopt^2} = ... =\frac{\partial^{q} t}{\partial \lopt^{q}} \,.
\]
In contrast to the case with only melonic interactions \cite{Bonzom:2012bg} where all equations are polynomial, in this case it is not obvious whether solutions to these systems of equations exist up to $q$ being the number of couplings.

Nevertheless, in explicit examples we do find multi-critical loci up to $q=m+1$ for $m\ge2$ melonic couplings, that is, even including the planar coupling $\tp$. 
This means that planar couplings enrich also the multi-critical structure of the branched-polymer regime which is a new insight to the best of our knowledge.

In the phase space of the models considered here, there are no intersections of multi-critical lines with transition lines. Accordingly, the case of critical exponents 
\[\label{eqref:multicritical}
\gamma \equiv 1 - \delta = \frac{p}{m+n+1} 
\]
as argued in \cite{AlvarezGaume:1992np} for matrix models and applied to tensor models \cite{Bonzom:2015gt} does not appear (in this relation, $m$ is the number of fine-tuned branched-polymer-type couplings, $n$ the number of fine-tuned planar couplings, and $p\le m$).

\subsection{Coupling melons and a necklace}

In this section we illustrate the phase structure of models with a planar coupling 
and a number of melonic couplings \eqref{eqref:DSEmelons+necklace} for a particular example. 
At least three couplings are necessary to investigate both the transition from a planar to a branched-polymer regime as well as interference with multi-criticality in the latter one.
%
The simplest non-trivial case is the tensor model with a quartic and sextic melonic and a quartic necklace interaction,
\[
\lop = -\gt + \tm 2 \gt^2 + \tm 3 \gt^3 \,.
\]
On the surface
\[
\tm{3}^{\pla}= \tm{3}^{\pla}(g,\tm2) = \frac{3}{4} \left(9 \tp - 2 \sqrt{ 3 \tp (9 \tp+\tm2)^2}\right) 
, \qquad \gt^\pla =\gt^\pla(\tp) =  \frac{2}{3 \sqrt{3 g}} \, ,
\]
we find $\gamma=-1/2$ critical behaviour
\[
\gt (\tp, \tm2, \tm3) =  \gt^\pla + \frac{4}{27 \tp \left(\tm2^\tra -\tm2\right)} \left(\tm3^\pla - \tm3 \right)
- \frac{32}{81 \sqrt[4]{3 \tp} \left(\tm2 - \tm2^\tra \right)^{5/2}} \left(\tm3^\pla - \tm3 \right)^{3/2}
+ \cO\left(\tm3^\pla - \tm3 \right)^{2} \, .
\]
The coefficient of the fractional-exponent expansion term already indicates, that this expansion breaks down for $\tm2 < \tm2^\tra$ where 
\be\label{eqref:transition-line}
\tm2^\tra = \tm2^\tra(\tp) = -3 \left(7 g - \sqrt{3 g}\right) \quad \textrm{and} \quad 
\tm3^\tra = \tm3^\tra(\tp) = -\frac{9}{4}g \left(3 - 8 \sqrt{3 g} \right) 
\ee
for $g>0$. Indeed, this is the solution we find for the line of phase transition between planar and branched-polymer regime.
We have checked the expansion around the transition line $(g,\tm2^\tra,\tm3^\tra)$, for example in $\tm3$,
\[
\gt (\tp, \tm2, \tm3) =  \gt^\pla - \frac{1}{9 \,3^{1/6} g^{3/2}}\left(\tm3^\tra - \tm3 \right)^{2/3} + \cO(\tm3^\tra - \tm3)
\]
yielding critical behaviour with exponent $\gamma = 1/3$.
In particular, using the Newton-polygon method  \cite[ch. VII.7]{Flajolet:2009wm} we find that there are no further multi-critical points on this line.

Critical surfaces on which $\gt$ has $\gamma=1/2$ critical behaviour are fourth roots of the equation
\bea
0 = P_\bp(g,\tm2,\tm3) & = & 3125 g^4 \tm3^2+7500 g^3 \tm2 \tm3^2-2 g^2 \left(2 \tm2^5+25 \tm2^3 \tm3-2775 \tm2^2 \tm3^2+500 \tm2 \tm3^2-4500 \tm3^3\right)\\
&& -4 g \left(2 \tm2^6+23 \tm2^4 \tm3-283 \tm2^3 \tm3^2+160 \tm2^2 \tm3^2-1260 \tm2 \tm3^3+16 \tm3^3 (25-108 \tm3)\right)\\
&& -\left(\tm2^2+4 \tm3\right)^2 \left(4 \tm2^3-\tm2^2+18 \tm2 \tm3+\tm3 (27 \tm3-4)\right)
\eea
whose solutions are hard to handle explicitly.
However, from the branching lines of the solutions we can find an explicit expression for the multi-critical line
\[\label{eqref:mcritical-line}
\begin{pmatrix}
\tp \\ \tm2 \\ \tm3 
\end{pmatrix}
= \frac{(5 \lambda +1)(25 \lambda ^2+118 \lambda +1) + \left(25 \lambda ^2-26 \lambda +1\right)^{\frac3 2}}{6\times72 \left(25 \lambda ^2+22 \lambda +1\right)^2} 
\begin{pmatrix}
72 \lambda\\ 72\\
625 \lambda ^3 +150 \lambda ^2  - 339 \lambda - 4 - (5 \lambda +4) \sqrt{\lambda -1} (25 \lambda -1)^{\frac3 2}
\end{pmatrix} \nonumber
\]
parametrized by $\lambda \le1/25$.
On this line we find $\gamma = 2/3$. 
The endpoint $(\tp , \tm2 , \tm3 ) = (1/{80}, 5/{16}, -5/{128})$ of this line is multi-critical with $\gamma = 3/4$ as can be seen from the expansion
\[
\gt\left(\frac1{80}, \frac5{16}-\tm2, -\frac5{128}\right) = \frac{16}{5}-\frac{32\ 2^{3/4}}{5 \sqrt{5}} \left({\frac{5}{16}-\tm2}\right)^{1/4} +\cO\left({\frac{5}{16}-\tm2}\right) \,.
\]

The surprising finding is that in this example the $\gamma = 3/4$ point lies close to but not on the phase-transition line \eqref{eqref:transition-line}. Moreover, we have checked that the transition line and the multi-critical line \eqref{eqref:mcritical-line} have no intersection. 
In particular, there is no critical point on the transition line with exponent $\gamma = 2/4 = 1/2$ as would follow from \eqref{eqref:multicritical}.
Therefore, the exponents argued for with matrix-model methods on the grounds of intersections of a multi-critical line with the line of phase transition \cite{AlvarezGaume:1992np} are not recovered here. 
For such a result it is necessary to  further prove that such intersections actually exist. Our example shows that there are cases where they do not exist.

\

The properties of the two-point function $\gt$ detailed in this example are common to solutions of the general type of   \eqref{eqref:DSEmelons+necklace}. We have checked combinations of various melonic interactions and, in particular, also type $3_\textrm{K}$ interactions instead of quartic necklaces.
A peculiar aspect of the type $3_\textrm{K}$ interaction $\bk$ is that it effectively contains already both planar and branched-polymer behaviour since the equations for the two-point function \eqref{eqref:DSE-K334} are of the form
\be\label{eqref:DSE-K334gen}
\gt = \nsgf \left(\tp \gt ^3\right)+ \tm3 \gt^3  + \tm6^2 \gt^6 \, ,
\ee
only with fixed balance of coupling $\tp=2\tk$, $\tm3=4\tk$ and $\tm6=6\tk$. Nevertheless, the tree-like part dominates in this case and $\gt$ has critical exponent $\gamma=1/2$ at the critical point.
This can be easily freed adding melonic interactions of type $3_\textrm{M}$ and $6_\textrm{M}$ such that then $\tm3=4\tk + \tmp3$ and $\tm6 = 6 \tk + \tmp6$. Then there is again the full phase-space with $\gamma=1/3$ phase transition between planar and branched-polymer regime.

\subsection{Two types of necklaces with independent couplings}
\label{subsec:TwoNecks}

There is an example of a tensor model with a planar and a branched-polymer phase which is conceptually even simpler: two or more types of necklace interactions.
It has been considered partially in \cite{Bonzom:2016to} for the case of quartic necklace interactions $\bnp2^{1i}$ with identical couplings $\tnp2^{12}=\tnp2^{13}$ and so forth. This results in geometries with branched-polymer behaviour as soon as there are more than one type of necklaces.

\begin{figure}
\centering
\includegraphics[width=7.5cm]{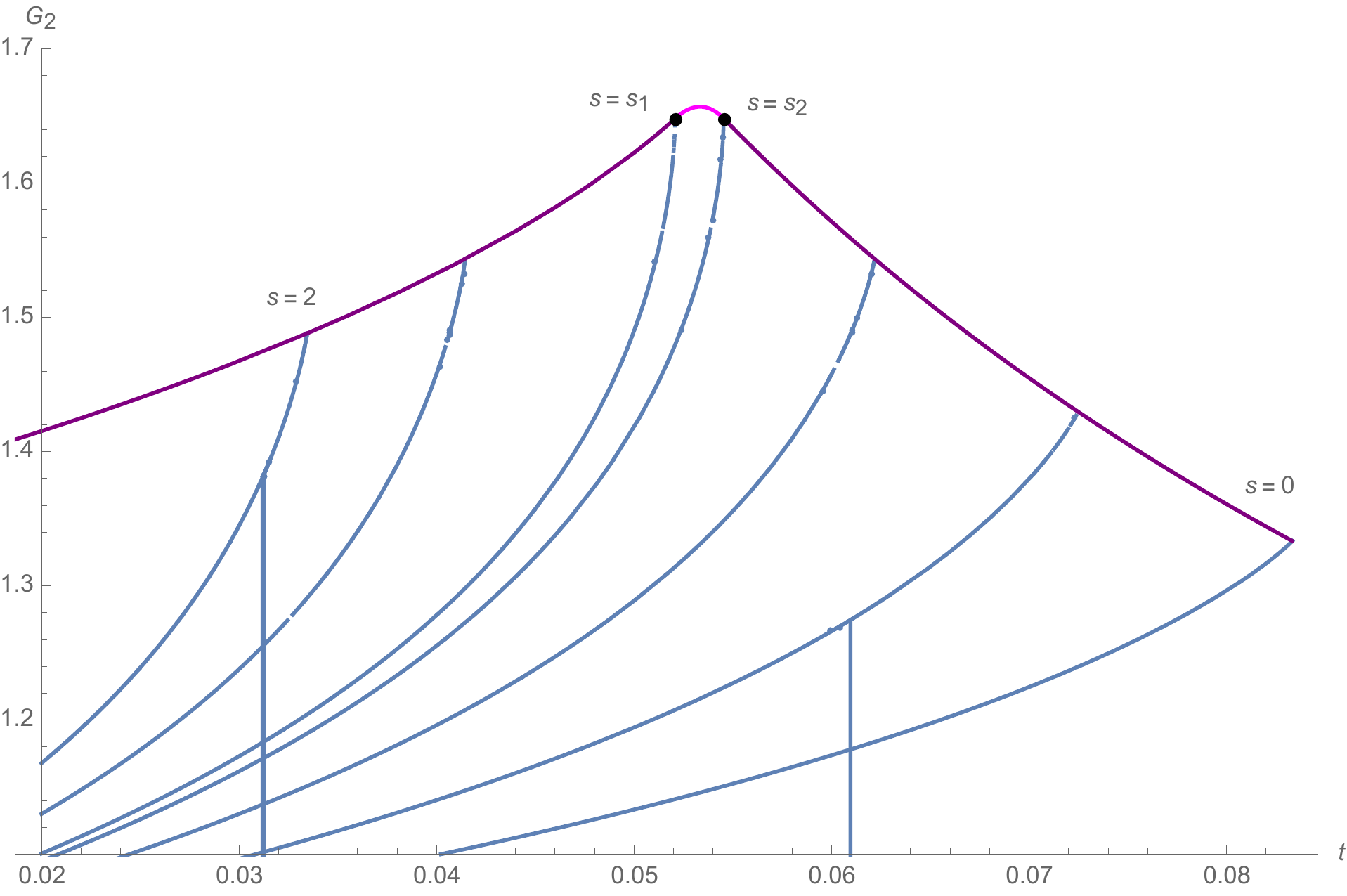}
\quad \quad
\includegraphics[width=7.5cm]{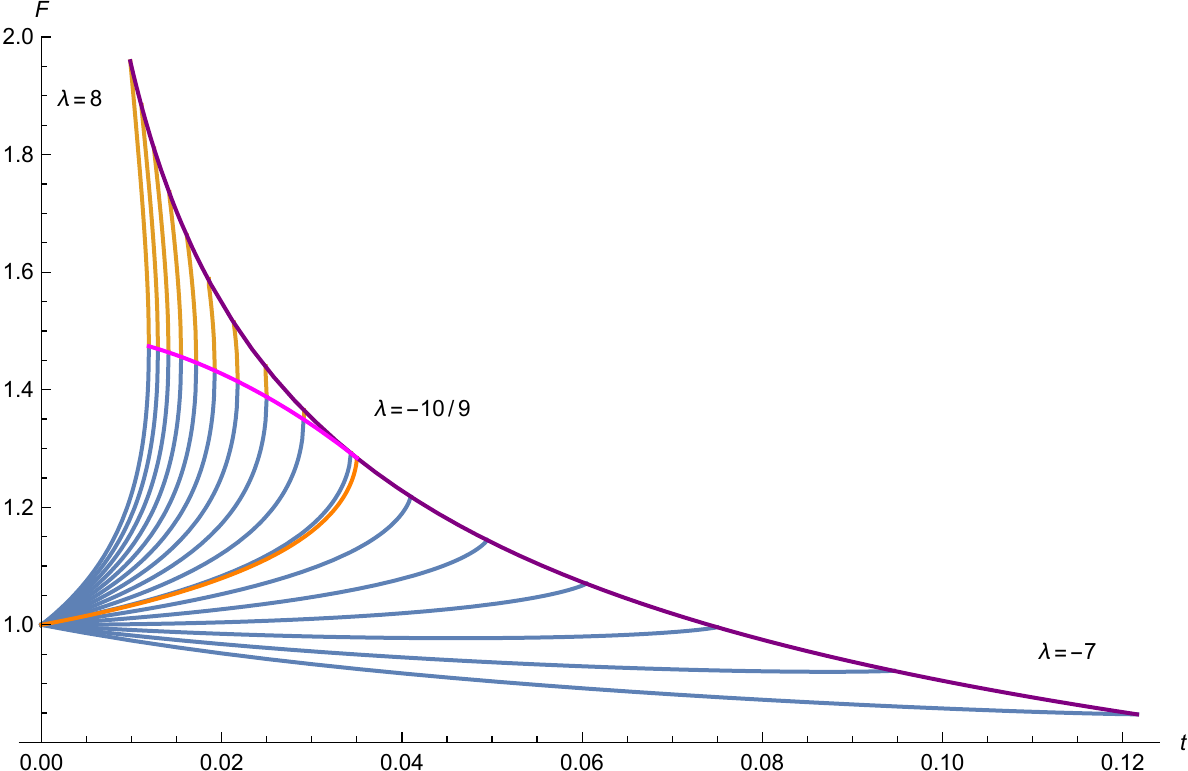}
\caption{Left: Two-point function $\gt(t,s)$ for two type $2_\textrm{N}$ quartic necklace interactions for various coupling ratios $s$ according to \eqref{eqref:DSE2necklaces}.The purple curve shows planar critical points valid for $s<s_{1}$ and $s>s_{2}$ while the magenta curve indicates the critical points of the branched-polymer regime for $s_{c,1}<s<s_{c,2}$.\\
Right: For comparison, two-point functions $\gt(t,\lambda)$ for type $3_\textrm{K}$ and type $3_\textrm{M}$ interactions in similar ratio of couplings $t=\tk$ and $\lambda = \frac{4\tk+\tmp3}{2\tk}$ 
according to \eqref{eqref:DSE-K334gen}.
The branched-polymer critical curve (magenta) extends here to arbitrary high $\lambda$, as does the planar critical curve for arbitrary small $\lambda$. The phase transition with $\gamma=1/3$ critical behaviour occurs at $\lambda=-10/9$.
\label{fig:necklaces}}
\end{figure}

Here we show that for independent couplings, \eg $\tnp2^{12},\tnp2^{13}$, there is only a very narrow regime of branched-polymer behaviour.
The two-point function $\gt$ (see {Fig.~\ref{fig:necklaces}}) is the solution to the equation
\[\label{eqref:DSE2necklaces}
\gt(\tnp2^{12},\tnp2^{13}) = \gt(t,s)
= \nsgf(t\, \gt^2) + \nsgf(s\, t\, \gt^2) -1
\]
where we use the relative coupling $s=\tnp2^{13}/\tnp2^{12}$ and $t=\tnp2^{12}$ (one can check that indeed all results are symmetric under $s\mapsto s^{-1}$).
Then we find that $\gt$ has the expansion 
\be\label{eqref:2necklace-expansion}
\gt(t,s) = \gt^\bp - 32 
\frac{{\left(2 s + 5 -5 \sqrt{2 s}\right)^{1/2} \left(5 s + 2 -5 \sqrt{2s}\right)^{1/2}}}{\left(s+1-\sqrt{2 s}\right)^{1/2}}
\left(t^\bp -t \right)^{1/2} +\cO\left(t^\bp -t \right)
\ee
typical for the branched-polymer regime only for $s_1 < s < s_2$ with
\be
s_1 = \frac{5}{4} \left(3-\sqrt{5}\right) \approx 0.955 \quad , \quad
s_2 = \frac{1}{5} \left(\sqrt{5}+3\right)
\,\approx\,1.047 \,
\ee
and the critical values are
\[
t^\bp = \frac{1}{32 \left(s+1-\sqrt{2 s}\right)} \quad, \quad
\gt^\bp = \frac{4 \sqrt{2 s}}{s+1 -\sqrt{2 s}}-8 \, .
\]
Again the expansion \eqref{eqref:2necklace-expansion} is consistent with the solutions $s_1,s_2$ for the transition points as the radiant in the numerator of the expansion coefficient is positive for $s_1<s<s_2$.%
\footnote{For $s<(3-\sqrt{5})/5\approx0,153$ and $s>5(3-\sqrt{5})/4\approx6.545$ the expansion \eqref{eqref:2necklace-expansion} is defined as well but describes singularities of non-physical ($\gt(0,s)\ne 1$) solutions of \eqref{eqref:DSE2necklaces}.}
At these phase-transition values of $s$ we find at critical coupling $t_1^\tra = \sqrt{5}+3/{96}$ and $t_2^\tra = 5/96$
\be
\gt(t,s_{1,2}) 
= \frac{4}{3} \left(\sqrt{5}-1\right) 
-48 (c_{1,2})^{1/3} \left(t_{1,2}^\tra -t\right)^{2/3}
+ \mathcal{O}\left(t_{1,2}^\tra -t\right)
\ee
with expansion coefficients $c_1=3 \left(17 \sqrt{5}-38\right)$ and $c_2= 6/\sqrt{5}-11/25$.

\section{Conclusions}

We analysed the critical behaviour of tensor models for all quartic and sextic interactions in dimension 3 and 4, and computed the unique scalings leading to a defined $1/N$ expansion, with non-trivial leading order, i.e. with infinitely many contributions at leading order.

Most models present a branched-polymer phase, a 2d pure gravity (planar) phase, and an intermediate phase with positive critical exponent, in which limit infinitely many smooth fractal spheres - baby universes - proliferate to form cacti with infinitely many branching points. 
While this phase structure has, to our knowledge, always been recovered from models with a matrix interaction and a polymerization perturbation, we provide a model generating cacti of spheres of one or another colour converging towards the smooth fractal Brownian sphere in the continuum, except for a very small region of the phase space for which the interference between the two colours is so strong it breaks down to a branched-polymer phase. 
At the boundaries of this region, the two colours compete to produce proliferating baby-universes. 
Furthermore, while in the context of matrix models, the polymerization terms were necessarily non-connected multi-trace interactions coming from higher order curvature perturbations to the Einstein-Hilbert action, in the present case, the considered models are all built from connected interactions, and can all be interpreted as $D$-dimensional discrete Einstein-Hilbert theories of gravity without higher order curvature perturbations.

We study explicitly the multi-critical behaviour of a random tensor model with one quartic matrix-like ``necklace" interaction and polymerization ``melonic" perturbations of orders 4 and 6 and see that instead of the entropy exponent $\gamma=2/4$ we would expect from \cite{AlvarezGaume:1992np}, we obtain the multi-critical branched-polymer exponent $\gamma=3/4$ at the end of the multi-critical branched-polymer line, the planar coupling playing the role of an higher order polymer interaction (we would not be able to reach an order-3 multi-critical branched-polymer exponent from only two quartic polymer interactions and a sextic one). We stress that all the models introduced can be understood on the level of discrete $D$-dimensional spaces dual to the Feynman stranded diagrams, so that all the critical and multi-critical phases involved have interpretations in terms of statistical systems coupled to random $D$-dimensional lattices with very specific adjacency conditions.

The results in this paper rely on a bijection between the Feynman stranded diagrams and embedded diagrams, which generalizes the so-called intermediate-field representation. Throughout this bijection, the link with matrix-models and two-dimensional discrete surfaces is made clear, and we therefore see that the non-planarity of the order 6 stranded interactions is replaced by a planar colouring rule on some cubic embedded vertices. In light of this, a more appropriate criterium for interactions candidates to generate more interesting continuum limits could be the lack of a planar embedded equivalent intermediate-field vertex from the bijection. 

\vspace{-0.5cm}
\section*{Acknowledgements}
\vspace{-0.3cm}

We would like to thank S. Dartois,  V. Bonzom, V. Rivasseau and J. Ben Geloun for useful discussions. 
The work leading to this publication was supported by the German Academic Exchange Service (DAAD) with funds from the German Federal Ministry of Education and Research (BMBF) and the People Programme (Marie Curie Actions) of the European Union's Seventh Framework Programme (FP7/2007-2013) under REA grant agreement n$^\circ$ 605728 (P.R.I.M.E. - Postdoctoral Researchers International Mobility Experience).

\appendix
\section{Correspondence between stranded Feynman graphs and embedded diagrams }
\label{appendix}

In this appendix, we summarize the one-to-one correspondence between stranded Feynman diagrams of random tensor models, and embedded diagrams, in the case of interactions of order 6. This correspondence generalizes the intermediate field representation for the quartic models \cite{Bonzom:2015gt,Gurau:2014kj,Delepouve:2014wk}. It was introduced in \cite{SWMaps} in a more combinatorial language for interactions of any order, and further developed in \cite{Lionni:17}.
See also \cite{Bonzom:2016to} for a pedagogic presentation.
Thanks to this correspondence, we can rewrite arbitrary random tensor models as multi-trace matrix models \cite{SWMaps, Lionni:17}.  

Let us describe the  
bijection $\Psi$ between stranded diagrams $\cG$ with sextic interactions $B_1, \ldots, B_n$, and embedded diagrams $\cM$ defined as follows:
\begin{itemize}
\item The diagrams have  white vertices as shown in  Tables~\ref{table:rank3-interactions} and~\ref{table:rank4-interactions}, corresponding to the interactions $B_1, \ldots, B_n$ under consideration. As represented in these tables, the edges attached to these white vertices carry specific sets of colours $\cI\subset \{1,\ldots,D\}$. The ordering of the edges around white vertices is fixed.
\item{The other extremities of the edges can only be attached to black vertices. There is no restriction on the number or the kind of edges attached to black vertices. The  edges around black vertices are cyclically ordered.}
\end{itemize}
If we take such an embedded diagram $\cM=\Psi(\cG)$ and delete all the edges whose sets of colours does not include the colour $i$, as well as all the isolated white vertices, we obtain a fatgraph $\cM_i$ equivalent to a ribbon-graph, whose faces are in one-to-one correspondence with the faces of colour $i$ of $\cG$. The interest of this correspondence is thus that we can use ribbon graphs and matrix models tools to identify the leading orders for random tensor models. 
\

 \be
\label{fig:ExBijApp}
\begin{array}{c}\includegraphics[scale=0.4]{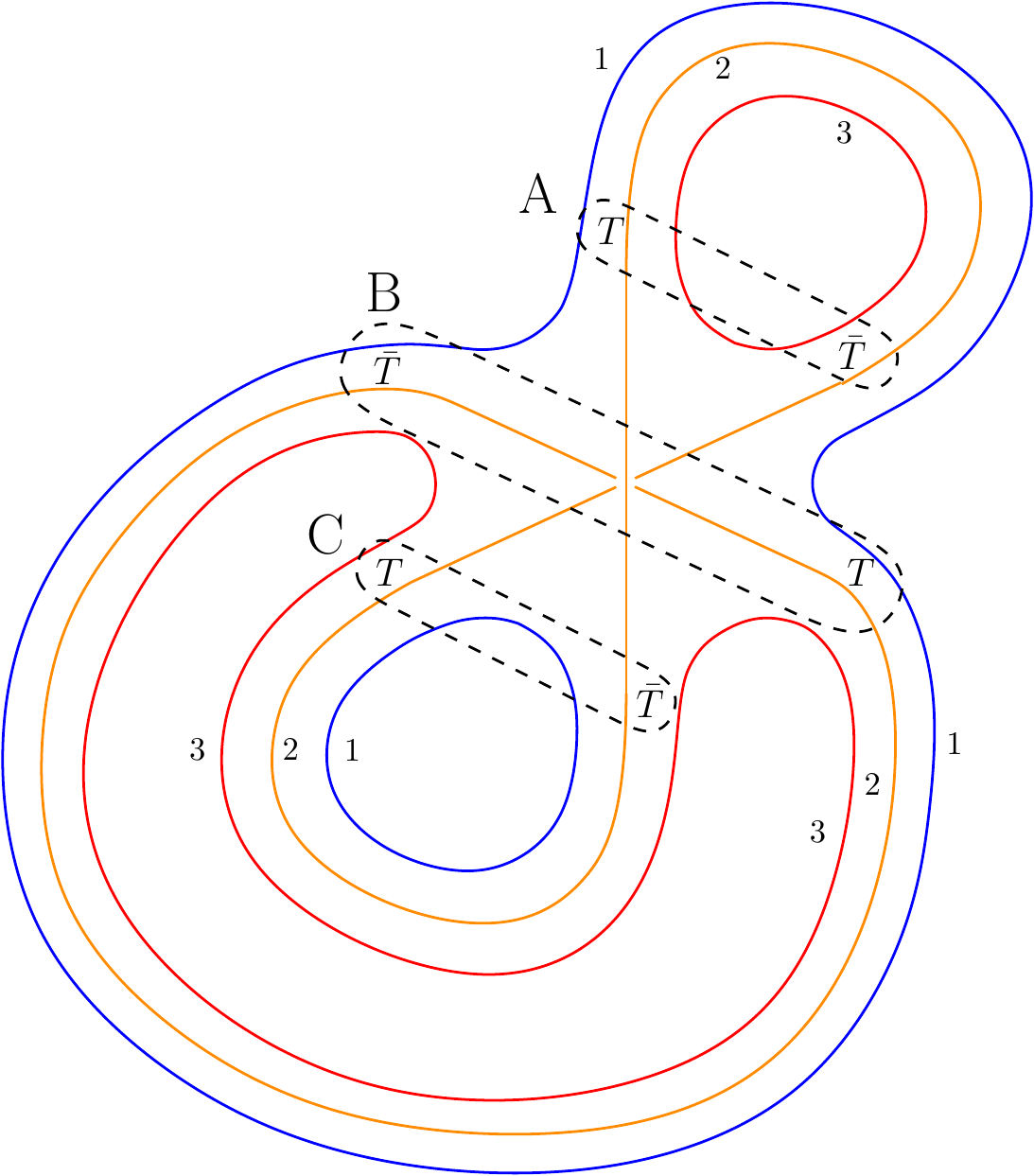}\end{array}
\hspace{1cm}\overset{\Psi}{\longrightarrow}\hspace{1cm}
\begin{array}{c}\includegraphics[scale=0.65]{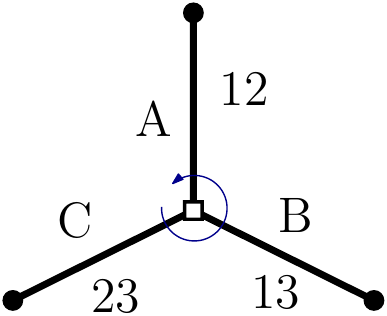}\end{array}
\ee

\

\noindent{\it From stranded graphs to embedded diagrams.} A choice has been made for every interaction, of a pairing of every tensor $T$ with a $\bar T$, and this defines the white vertices in Tables~\ref{table:rank3-interactions} and~\ref{table:rank4-interactions}. To indicate the pairing in the tables, we have used the graphical convention (\ref{fig:Pairings}). The pairs $(T, \bar T)$ in the stranded graph $\cG$ correspond to the edges around the white vertices in $\Psi(\cG)$. By following alternatively the propagators and the pairs $(T, \bar T)$  in the stranded graph $\cG$, we partition the pairs $(T, \bar T)$ into  cyclically ordered groups (the propagators are crossed from the $\bar T$ to the $T$). All the edges corresponding to pairs $(T, \bar T)$ in the same group are connected to a common black vertex in $\Psi(\cG)$. These edges are cyclically ordered, due to the ordering of the pairs in $(T, \bar T)$ in the group. This defines the diagram $\Psi(\cG)$. 

\

If as a convention the tensor $T$ (resp.~$\bar T$) leaves on the left (resp.~right) of the edge, when seen from the white vertex, then  the propagators in $\cG$ correspond to the corners around black vertices in $\Psi(\cG)$, and they are followed counterclockwise when the propagators are crossed from the $\bar T$ to the $T$ in $\cG$. The white vertices in Tables~\ref{table:rank3-interactions} and~\ref{table:rank4-interactions} render the internal coloured structure of the stranded interaction : the corners around white vertices in $\cM_i$ correspond to strands of colour $i$ between tensors inside the interaction in $\cG$. Moreover, if a colour $i$ is not included in the set of colours of an edge, it means that an edge of colour $i$ goes between the two tensors of the corresponding pair $(T, \bar T)$.  Thus the corners around black vertices in $\cM_i$ also correspond to strands of colour $i$ between tensors in $\cG$. This explains why the faces around $\cM_i$ correspond to faces of colour $i $ in $\cG$.

 \be
\label{fig:ExBijApp2}
\begin{array}{c}\includegraphics[scale=0.42]{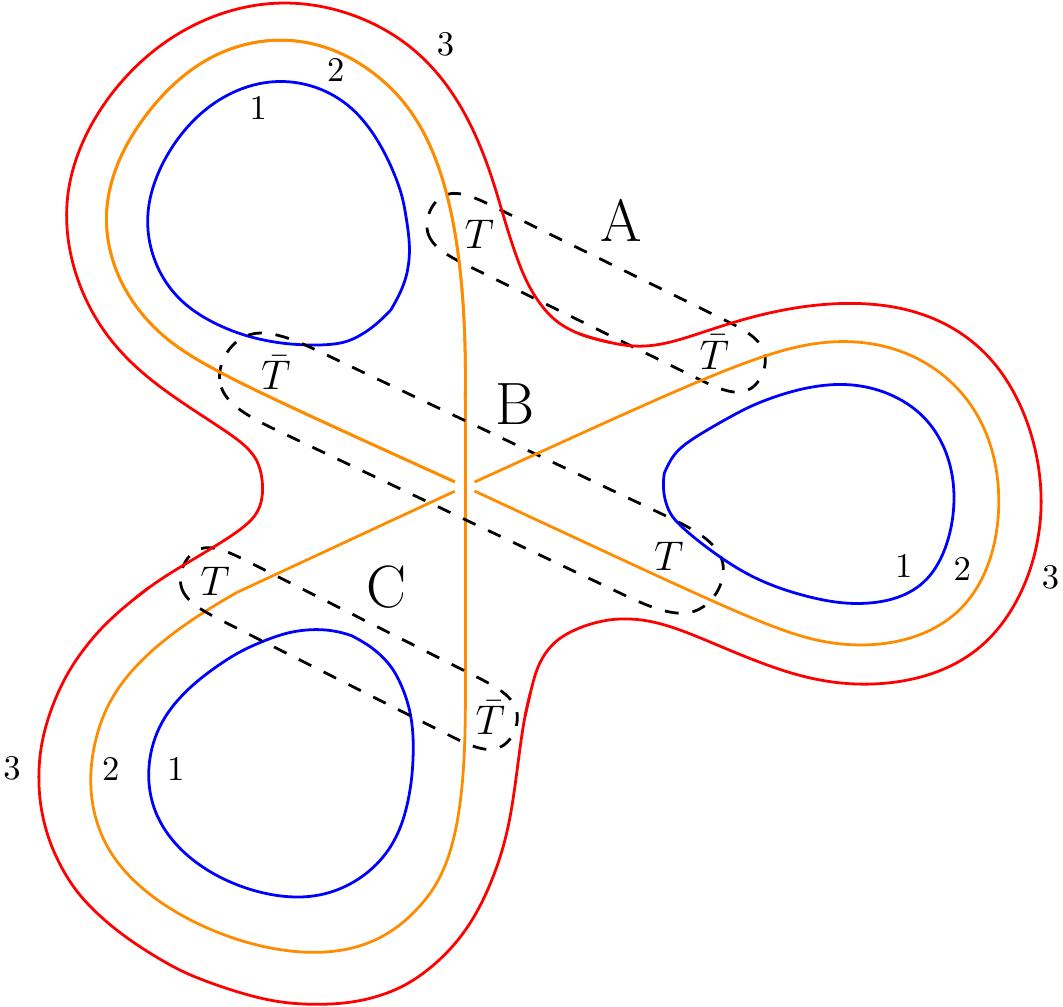}\end{array}
\hspace{0.3cm}\overset{\Psi}{\longrightarrow}\hspace{0.1cm}
\begin{array}{c}\includegraphics[scale=0.65]{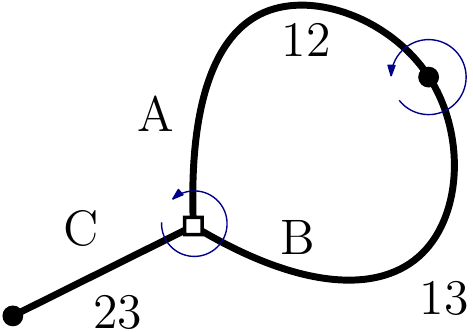}\end{array}
\ee

\noindent{\it From embedded diagrams to stranded graphs.} Given an embedded diagram $\cM$, replace each white vertex by the stranded representation, by matching the pairs $(T, \bar T)$ with the corresponding edges. Be sure to respect the convention that the $T$ (resp.~$\bar T$) are on the left (resp.~right) of the edge when seen from the white vertex. For every corner around a black vertex in $\cM$, add a propagator between the corresponding $T$ and $\bar T$ in the stranded graph.

\bibliographystyle{JHEP}
\bibliography{main}

\end{document}